\documentclass[]{aa}
\usepackage{txfonts}
\usepackage{graphicx} 
\numberwithin{equation}{section}
\usepackage{hyperref}
\usepackage{natbib}
\usepackage{subcaption}
\usepackage{color}
\usepackage[normalem]{ulem}
\usepackage{gensymb}

\newcommand{\Ha}{\mathcal H} 
\newcommand{\GravC}{\mathcal G} 

\newcommand{\E}{E} 
\newcommand{\M}{\ell} 
\newcommand{\I}{I} 
\newcommand{\VARPI}{\varpi} 
\newcommand{\LAMBDA}{\lambda} 
\newcommand{\GAMMA}{\gamma}

\newcommand{\LAMBDONA}{\Lambda}
\newcommand{\GAMMONA}{\Gamma}

\newcommand{\KAPPONA}{\mathcal K}
\newcommand{\KAPPA}{\kappa}
\newcommand{\THETA}{\theta}
\newcommand{\THETAONA}{\Theta}

\newcommand{\PSI}{\psi}
\newcommand{\PSIONA}{\Psi}
\newcommand{\DELTAGAMMA}{\delta\GAMMA}
\newcommand{\THETAP}{\theta'}
\newcommand{\OMEGONA}{\Omega}
\newcommand{\ANGMOM}{\mathcal L}

\newcommand{\days}{{~\mathrm{days}}}

\begin{document}
\font\courier=pcrr at 10pt
\font\codefont=pcrr

\title{The role of dissipative evolution for three-planet, near-resonant extrasolar systems}
\author{
Gabriele Pichierri\inst{1},
Konstantin Batygin\inst{2}
\and
Alessandro Morbidelli\inst{1}
          }

\institute{
Universit\'e C\^ote d'Azur, Observatoire de la C\^ote d'Azur, CNRS, Laboratoire Lagrange, France\\
\email{gabriele.pichierri@oca.eu} 
\and 
Division of Geological and Planetary Sciences, California Institute of Technology, 1200 E. California Blvd., Pasadena, CA 91125, USA
             }
\authorrunning{Pichierri, Batygin, Morbidelli}
\date{\today}
\abstract{
Early dynamical evolution of close-in planetary systems is shaped by an intricate combination of planetary gravitational interactions, orbital migration, and dissipative effects. While the process of convergent orbital migration is expected to routinely yield resonant planetary systems, previous analyses have shown that the semi-major axes of initially resonant pairs of planets will gradually diverge under the influence of long-term energy damping, producing an overabundance of planetary period ratios in slight excess of exact commensurability. While this feature is clearly evident in the orbital distribution of close-in extrasolar planets, the existing theoretical picture is limited to the specific case of the planetary three-body problem. In this study, we generalise the framework of dissipative divergence of resonant orbits to multi-resonant chains, and apply our results to the current observational census of well-characterised three-planet systems. Focusing on the 2:1 and 3:2 commensurabilities, we identify three 3-planet systems, whose current orbital architecture is consistent with an evolutionary history wherein convergent migration first locks the planets into a multi-resonant configuration and subsequent dissipation repels the orbits away from exact commensurability. Nevertheless, we find that the architecture of the overall sample of multi planetary systems is incompatible with this simple scenario, suggesting that additional physical mechanisms must play a dominant role during the early stages of planetary systems' dynamical evolution.
}
\maketitle

\section{Introduction}\label{sec:Introduction}

\begin{figure*}[!ht]
\centering
\begin{subfigure}[b]{0.45 \textwidth}
\centering
\includegraphics[scale=0.6
]{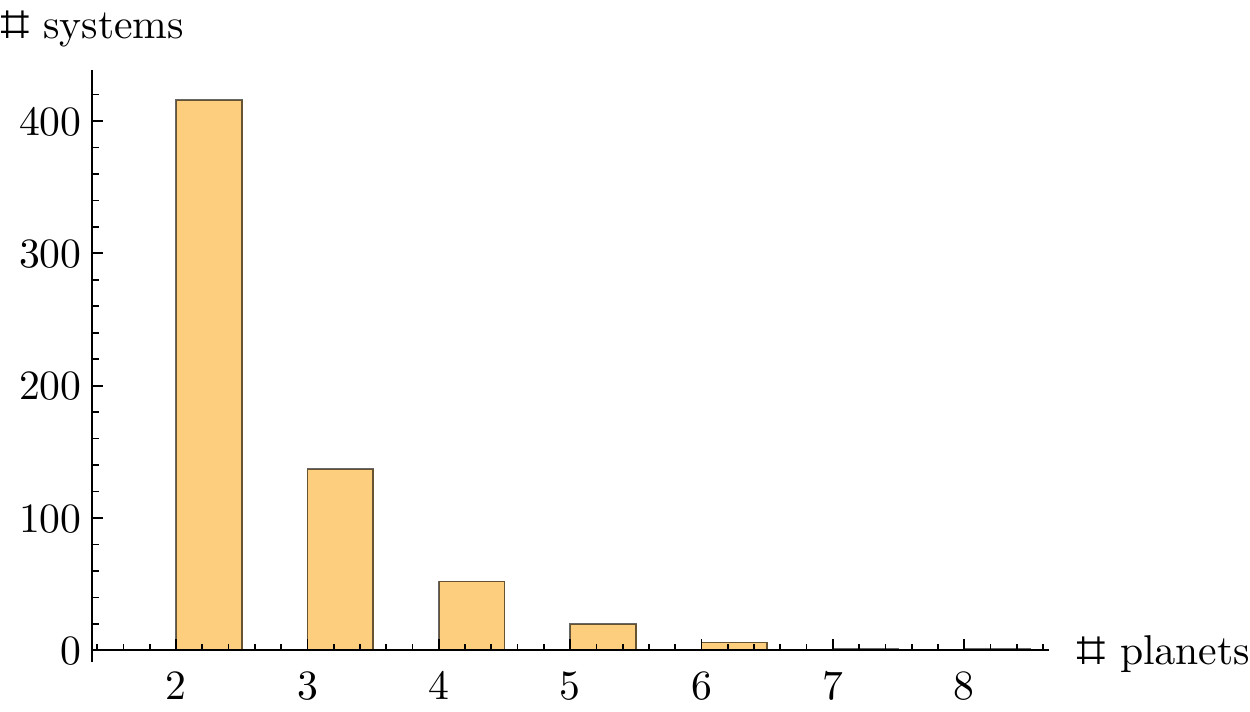}

\caption{Histogram of multi-planetary system by number of planets in each system.}
\end{subfigure}
\hfill
\begin{subfigure}[b]{0.45 \textwidth}
\centering
\includegraphics[scale=0.6
]{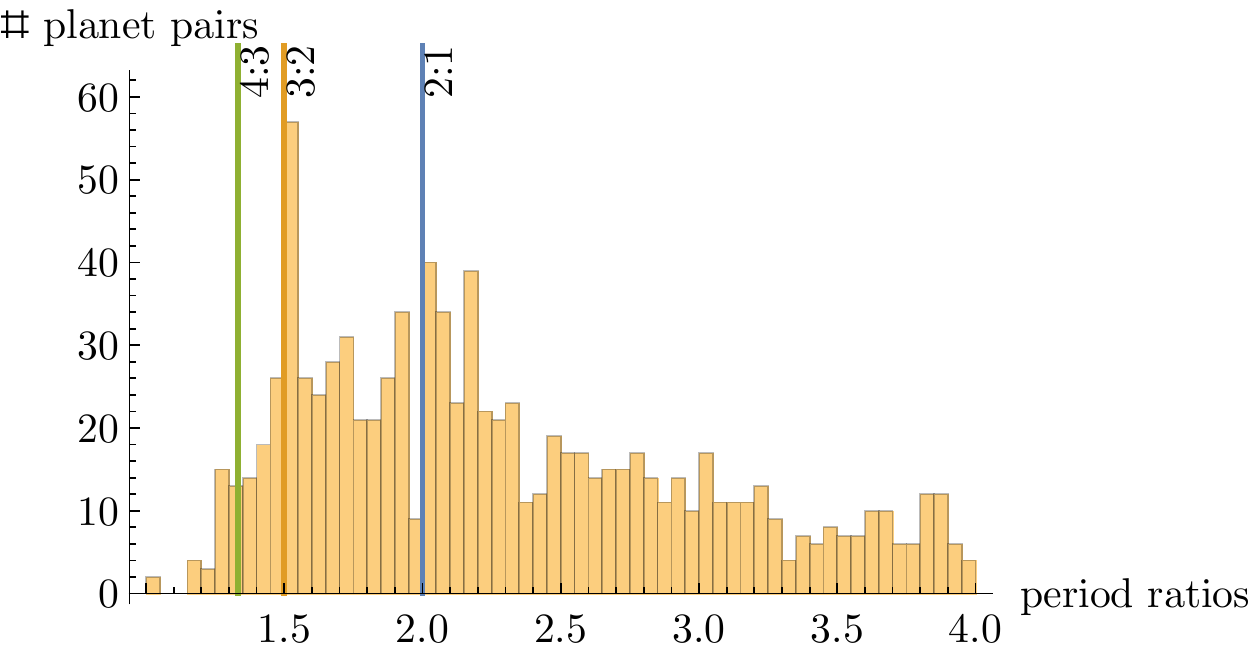}
\caption{Distribution of period ratios of neighbouring planets in exoplanetary systems}
\end{subfigure}



\caption%
{Observations of planet-hosting stars reveal that multi-planetary systems are not rare, hosting over 1600 confirmed planets (panel (a)). The period ratio distribution of neighbouring planets is shown in panel (b). One can observe an overall broad distribution as well as a number of peaks slightly \emph{wide} of resonant ratios, especially the 2:1 and 3:2 commensurabilities. Data was obtained from the Nasa Exoplanet Archive \url{https://exoplanetarchive.ipac.caltech.edu/}.
}
\label{fig:ExoplanetsHistograms} 
\end{figure*}

The search for exoplanets in recent years has uncovered a multitude of planetary systems, the study of which is the key to an understanding of planetary formation and evolution. Currently, the exoplanet population is dominated by Kepler's transit detections, making the planetary physical radii and orbital periods the better constrained parameters of the sample. Concerning the first aspect, much work has been done recently to understand how photoevaporation sculpts the physical radii of planets (\citealt{2017AJ....154..109F} and references therein). In this work we address the second, complementary problem of the orbital period distribution.
One of the most notable aspects of the Kepler data is that the distribution of the period ratios of neighbouring planets in multi-planets systems shows two seemingly conflicting features: on the one hand, it appears relatively broad and smooth, without any single, unmistakably emerging feature; on the other hand, a slight preference for near-resonant configurations is evident upon close examination. In fact, it is often pointed out that there is a lack of planet pairs in correspondence with period ratios very close to low-integer ratios, and a definite excess just wide of these values, especially the 2:1 and 3:2, see Figure \ref{fig:ExoplanetsHistograms}.

Numerical simulations show that compact chains of mean motion resonances are a common outcome of slow, convergent orbital transport of planets within protoplanetary disks. Although details of disk-driven migration remain an active topic of research, it is clear that such a process should play some role in the dynamical history of planetary systems. For example, it is not easy to envision a formation narrative which does not require convergent migration for systems such as Trappist-1, a star famously hosting 7 planets with period ratios very close to small integer ratios (\cite{2016Natur.533..221G,2017Natur.542..456G}, \citealt{2017NatAs...1E.129L}). Indeed, \cite{2017Natur.542..456G} performed $N$-body integrations with the orbital fits as initial conditions and these went unstable over timescales 10,000 times shorter than the estimated age of the system; in contrast, \cite{2017ApJ...840L..19T} remarked that if an initial condition which results from capture into resonance through migration is chosen, then the system is stable over timescales two orders of magnitude longer then the ones found in \cite{2017Natur.542..456G}. They also note that the addition of tidal eccentricity damping should help maintain the evolution stable over the system's age. Other good examples of systems necessarily sculpted by migration are the four sub-Neptune planets of Kepler-223 \citep{2016Natur.533..509M} and the now-classic example of Laplace-like resonance in GJ-876 \citep{2010ApJ...719..890R,2015AJ....149..167B}.

In light of the fact that convergent migration should lock planets into mean motion commensurability, how can we explain the lack of planets with exactly resonant period ratios and the excess just wide of them?
Analytical models of resonance do predict that a pair of planets in a first order mean motion resonance need not satisfy the exact resonance condition $a_1/a_2=((k-1)/k)^{2/3}$ (where $a_1$ and $a_2$ are the semi-major axes of the inner and outer planet, respectively, and $k$ is a positive integer), but they can reside wide of resonance while the resonant angles are still librating. This divergence of the resonant equilibrium configurations happens at vanishingly low eccentricities and is linked to a fast precession of the perihelia, which is well understood analytically. However some Kepler systems are so wide of resonance that, after the resonant configuration is attained and the disk of gas is dissipated, an auxiliary mechanism might need to be invoked which actively drives these planets farther away from the exact resonance. As we will see in Section \ref{sec:TestingScenario}, observations show that a significant fraction of nearly resonant systems lie up to 50 times wider from the resonance than the typical resonant width, and at lower eccentricities than are expected for such planets captured in resonance via migration in protoplanetary disks. These observations can potentially be interpreted as evidence for dissipative processes acting on the planetary systems after the disk phase.

\cite{2010MNRAS.405..573P} considered the specific case of the K-dwarf HD 40307, which hosts\footnote{
Note that since the publication of the aforementioned paper, more planets have been observed in the same system, including two confirmed planets HD 40307 f and HD 40307 g. For this reason, we will not consider this system in the current work, although we draw inspiration from the analysis of \cite{2010MNRAS.405..573P}.
} 3 hot super-Earths/mini-Neptunes with both pairs wide of the 2:1 mean motion resonance, and planetary masses obtained with Radial Velocity. They showed that as tidal interaction between the planets and the star reduces the eccentricities, the system maintains the libration of the resonant angles even when the period ratios are considerably far away from exact commensurability. 
Subsequently, \cite{2013AJ....145....1B}, \cite{2012ApJ...756L..11L} and \cite{2014A&A...570L...7D} showed that two planets in mean motion resonance repel each other as energy is lost during tidal evolution. They thus proposed this as a viable mechanism to explain the observed distribution of period ratios in exoplanetary systems.
Note that, for two planets, this repulsion can be easily understood if one considers that any process that dissipates the energy, $E\propto -m_1/a_1-m_2/a_2$, and at the same time conserves angular momentum, $L\propto m_1\sqrt{a_1} + m_2\sqrt{a_2}$, should give rise to such an evolution. Indeed, this study applies to \emph{any} dissipative evolution that maintains constant the angular momentum, not just tidal dissipation, and not just resonant coupling \citep{2012A&A...546A..71D}. 

Thanks to these studies, the case of two planet system is well understood. However the data also contains numerous systems of more than two planets (Figure \ref{fig:ExoplanetsHistograms}). Accordingly, in this paper we aim to expand the study to detected extrasolar systems of three planets.
More specifically, we envision the following scenario for the formation and evolution of these planetary systems. 
First, three planets are embedded in the protoplanetary disk in which they formed; they interact with the disk, which ultimately results in a resonant capture. Then, after the disk is slowly depleted, the dissipative effects mentioned above are introduced, leading to orbital divergence.

Naturally this is a simplified and idealised scenario. In reality, we still do not know with enough accuracy the final configuration obtained by multi-systems migrating in a disk of gas. 
One approach towards a better approximation would be to perform full hydrodynamic simulations of planets immersed in their protoplanetary disk accounting for various disk parameters (such as disk surface density, turbulence, opacity, etc.). This approach would however be very expensive computationally.
Moreover, to date we have virtually no direct observations of the specific physical processes acting during planet formation and evolution in the early epochs of the disk phase, so these simulations, no matter how exhaustive in terms of the implementation of the plausible physics, cannot yet be directly constrained by the available data.
In any case, the fact that slow convergent orbital transport strongly favours resonant captured states is well supported both analytically and numerically, as well as by the specific observations of multi-planets systems mentioned above. 

In this paper we focus on slow convergent Type-I migration in a disk of gas, and adopt simple synthetic analytical formul\ae\ for the work and the torque generated by the disk on the planets (\citealt{2006A&A...450..833C,2008A&A...482..677C}); the requirement that exact prescriptions for the interaction between the planets and the disk be implemented will be relaxed, invoking the aforementioned arguments favouring the plausibility of mean motion resonant capture. A similar reasoning can be applied for the post-disk phase. In order to simulate the dissipative forces that can act on the planetary system, we will implement tidal dissipation. Of course, the tidal parameters for these planets are not known (as we do not yet have a precise understanding of the interior structure of these bodies or the specific physical mechanisms that dominate the dissipation), which would pose additional questions concerning for example the timescales over which this type of dissipation takes place. However, the specific choice of tidal dissipation is only one possible example of a process such that $\dot E<0$ and $\dot\ANGMOM=0$. 
We conclude that our specific implementation of Type-I migration and tidal dissipation after the disk removal are therefore not restrictive, which makes our results generalisable to any other equivalent processes. In the light of these considerations, we ask the following question: assuming that planets are captured into resonance and undergo dissipative evolution after the disk phase, is it possible to reproduce the observed orbital configuration of real exoplanetary systems which reside close to resonance? In other words: are the aforementioned physical processes compatible with the distribution of near-resonant period ratios that emerges from available data?

\subsection{Methods and physical setup}
In order to answer this question, we examined the NASA Exoplanet Archive (\url{https://exoplanetarchive.ipac.caltech.edu}) and selected confirmed 3-planet systems for which both planet pairs lie close to a first order mean motion resonance, in particular the 2:1 and 3:2 resonances, as these seem to be the most common in the Kepler data.
Our aim is to analyse these systems' orbital parameters and to evaluate quantitatively how close they are to a multi-resonant chain, which would be indicative of a dynamical history characterised by the physical processes described above.

Evidence suggesting that planets around Trappist-1 and Kepler-223 truly reside in a resonant configuration has recently been marshalled from the observed libration of the three-body Laplace angles. To this end, recall that if two neighbouring pairs of planets in a multi-planet systems are in the $k^{(in)}$:$(k^{(in)}-1)$ and $k^{(out)}$:$(k^{(out)}-1)$ resonances respectively (so that the resonant angles $k^{(in)}\LAMBDA_2-(k^{(in)}-1)\LAMBDA_1-\VARPI_2$ and $k^{(out)}\LAMBDA_3 -(k^{(out)}-1)\LAMBDA_2-\VARPI_2$ are librating), then the Laplance angle $\varphi_L=(k^{(in)}-1)\LAMBDA_1 - (k^{(in)} + k^{(out)}-1)\LAMBDA_2 + k^{(out)}\LAMBDA_3$ will be automatically librating as well. The advantage of examining this three-body angle over the two-body resonant angles is that the latter contain the longitudes of the pericenters $\VARPI$, whose precession rates are poorly constrained by the data, while the former only includes the mean longitudes $\LAMBDA$ whose derivatives in time are directly deduced by the transit observations.
However, solutions for which the resonant angles were originally in libration around a resonant equilibrium point can become circulating when the eccentricity of the equilibrium point becomes small enough under the effect of tidal damping \citep{2015A&A...579A.128D}, and, similarly, even a small distance from the equilibrium point could be responsible for breaking the libration of the three-body Laplace angle when the equilibrium eccentricity becomes small enough. 
Therefore, even if such circulations of the angles were observed, this would not be in disagreement with the envisioned scenario of resonant capture and subsequent dissipative evolution. In other words, the libration of the Laplace angle is a sufficient, but not necessary condition for past resonant capture in a chain of first-order resonances.

We therefore perform here a different analysis of the observed data, where we do not attempt to verify that a given system resides formally in resonance at the present day, but instead we evaluate the distance of a system from the considered resonance chain and the probability that this proximity is due to mere chance.   
In order to do this, we look for resonant solutions that provide the closest match to the observed planetary orbital configurations, that is the semi-major axis ratios.
It is worth anticipating here the following important point. As it will be clear later (see Section \ref{sec:RealSystemsStudy.subsec.Analytical}), in the case of only two resonant planets residing wide of resonance it is \emph{always} possible to find a resonant configuration which matches the observed data. This is because the eccentricities of these planets are at the present day not well constrained observationally, making the total orbital momentum of the system $\ANGMOM$ a free parameter: it is therefore always possible to find a value of $\ANGMOM$ that reproduces the observed $a_2/a_1$ with resonance-locked orbits. However, this is \emph{not} the case for systems of three planets, since we still have only one free parameter $\ANGMOM$ (whose value is linked to the initial captured state, not constrained observationally) but \emph{two} observables, that is the two pairs' semi-major axis ratios.

As detailed below, we carried out our study of finding orbital configurations that match the observed data using both an analytical and a numerical approach. The semi-major axes of the planets may be inferred from the orbital periods once the stellar mass is known, however this quantity is not yet well constrained in all cases. Nonetheless, all we will be interested in will be the semi-major axes \emph{ratios} $a_2/a_1$ and $a_3/a_2$, which can be obtained without any knowledge of the mass of the star directly from the period ratios and using Kepler's third law. This is tantamount to renormalising all separations by some arbitrary length, which does not affect the underlying physics since the dynamics only depends on the ratios of the semi-major axes and not on their individual values (only the timescale of the evolution does).

For the purposes of this study, we can limit ourselves to an analysis to first order in the planetary eccentricities. Indeed, the eccentricities that are expected for planets that have been captured into mean motion resonance by slow convergent migration in a disk are of order $\sqrt{\tau_e/\tau_a}\sim h$, where $\tau_a$ and $\tau_e$ are the timescales of migration and eccentricity damping respectively, and $h=H/r\sim 0.05$ is the aspect ratio of the disk (\citealt{2014AJ....147...32G}, \citealt{2018CeMDA.130...54P}). Since disks with high aspect ratios are not expected, the limit of small eccentricity is justified, and even more so in the phase of dissipative tidal evolution, which further damps the eccentricities. 
Moreover, given that these are transiting planets, and that during the disk phase any mutual inclination of the planets would be damped out, we assume coplanar orbits for simplicity.
Another useful piece of information which is available to us is the radii of the planets. This could in principle be used to infer the planetary masses (e.g.\ \citealt{2013ApJ...772...74W}). However the radius-mass relationship in Kepler planets is marked by extreme scatter \citep{2013ApJ...768...14W}, and we therefore choose to keep the planetary masses as a free parameter. More specifically, we are only interested in the mass ratios $m_1/m_2$ and $m_2/m_3$, since, as we will show, they are the only dynamically significant quantities that can affect the values of the semi-major axis ratios (see also the Appendix \ref{appendix:ReducedHamiltonian}). 

The remainder of this paper is organised as follows. In Section \ref{sec:PlanetaryHamiltonian} we obtain an analytical model for three planets in a chain of first order mean motion resonances, valid in the limit of small eccentricities. With this analytical model, we find the stable resonant configurations and map them in terms of the orbital elements. Finally we obtain an analytical confirmation of resonant repulsion for three-planets systems undergoing dissipation. In Section \ref{sec:TestingScenario} we detail our study, employing both analytical and numerical methods. We select systems of three planets near mean motion resonances, focusing on the 2:1 and 3:2 resonances, and we analyse their orbital configuration using the available data in order to evaluate if they are consistent with the process of resonant capture and subsequent dissipative evolution. We present our results in Section \ref{sec:Results} and we finally conclude by discussing their significance in Section \ref{sec:Conclusions}.

\section{Planetary Hamiltonian}\label{sec:PlanetaryHamiltonian}
The Hamiltonian of two resonant planets in the limit of low eccentricities has been studied extensively in the literature (e.g.\ \cite{2013A&A...556A..28B}, and references therein). Collectively these studies have pointed out that even if both planets are massive and to first order in eccentricity it is possible to reduce the problem to an Hamiltonian that is analogous to the well-known Hamiltonian of the restricted, circular three-body problem of a massless particle in resonance with a massive unperturbed body. In particular, such a Hamiltonian is integrable and is equivalent to the so-called second fundamental model for resonance \citep{1983CeMec..30..197H}. This is, however, not the case for three planets. Nonetheless, it is useful to extend an analytical description of the resonant dynamics at low amplitude of libration of the resonant angles in the case of three planets orbiting a star. In this section, we introduce the Hamiltonian of the system, derive curves representing the loci of its stable equilibrium points, and show how these can provide a description of a system along the dissipative evolution. We will apply this model to real Kepler system in Section \ref{sec:RealSystemsStudy.subsec.Analytical}.

Consider three planets of masses $m_1$, $m_2$ and $m_3$, orbiting around a star of mass $M_*$ in a canonical heliocentric reference frame \citep{1892mnmc.book.....P}. Indices 1, 2 and 3 will refer to the inner, middle and outer planet respectively. 
As usual, we consider the planetary Hamiltonian, which we write as
\begin{equation}\label{eq:FullHamiltonian}
\Ha = \Ha_{kepl} + \Ha_{pert},
\end{equation}
where the keplerian part is given by
\begin{equation}\label{eq:KeplerianPartInOrbitalElements}
\Ha_{kepl} = -\frac{\GravC M_* m_1}{2a_1} - \frac{\GravC M_* m_2}{2a_2} - \frac{\GravC M_* m_3}{2a_3},
\end{equation}
and describes the (integrable) motion of the three planets due to their interaction with the star, to which the small perturbation $\Ha_{pert}$ is added, which includes all the mutual interactions between the planets. We now assume that the inner pair of planets is close to a $k^{in}$:$(k^{in}-1)$ mean motion resonance, and that the outer pair of planets is close to a $k^{out}$:$(k^{out}-1)$ mean motion resonance, where $k^{in},~k^{out} > 1$ are two positive integers. In other words, we assume the resonance conditions $n_1/n_2\simeq k^{in}/(k^{in}-1)$, $n_2/n_3\simeq k^{out}/(k^{out}-1)$, where for each planet $n=\sqrt{\GravC(M_*+m) a^{-3}}$ is the mean motion. Since we are interested in the resonant interaction between the planets only, we will average the Hamiltonian over the fast evolving angles so that only combinations of the resonant angles $k^{in}\LAMBDA_2-(k^{in}-1)\LAMBDA_1 - \VARPI_1$, $k^{in}\LAMBDA_2-(k^{in}-1)\LAMBDA_1 - \VARPI_2$, $k^{out}\LAMBDA_3-(k^{out}-1)\LAMBDA_2 - \VARPI_2$, and $k^{out}\LAMBDA_3-(k^{out}-1)\LAMBDA_2 - \VARPI_3$ remain in the Hamiltonian, where $\LAMBDA$ is the mean longitude of a planet, and $\VARPI$ is its longitude of the periastron.

The resonant perturbing Hamiltonian expanded to first order in the eccentricities reads
\begin{equation}\label{eq:ResonantHamiltonianInOrbitalElements}
\begin{split}
\Ha_{res} 	&=	-\frac{\GravC m_1 m_2}{a_2}\left(f_{res}^{(1,in)}e_1 \cos\big(k^{in}\LAMBDA_2-(k^{in}-1)\LAMBDA_1 - \VARPI_1\big) + \right.\\ 
			&\quad\quad\quad \left.+f_{res}^{(2,in)}e_2 \cos\big(k^{in}\LAMBDA_2-(k^{in}-1)\LAMBDA_1 - \VARPI_2\big)
			 \right) + \\
			 &-\frac{\GravC m_2 m_3}{a_3}\left(f_{res}^{(1,out)}e_2 \cos\big(k^{out}\LAMBDA_3-(k^{out}-1)\LAMBDA_2 - \VARPI_2\big) + \right.\\ 
			&\quad\quad\quad \left.+f_{res}^{(2,out)}e_3 \cos\big(k^{out}\LAMBDA_3-(k^{out}-1)\LAMBDA_2 - \VARPI_3\big)
			 \right),
\end{split}
\end{equation}
where the orbital elements are constructed from heliocentric positions and barycentric velocities \citep{1892mnmc.book.....P}.
The coefficients $f_{res}$ are typically of order unity, and it is straightforward to determine the strength of each resonant harmonic, and incorporate direct and indirect terms. They depend (weakly) on the semi-major axis ratios, and their expressions may be found in \cite{2000ssd..book.....M}.
We therefore write the Hamiltonian after the averaging procedure as 
\begin{equation}
\bar\Ha = \Ha_{kepl} + \Ha_{res} + \mathcal O(e^2,\I^2),
\end{equation}
and then drop the higher order terms.
Note that terms that describe the mutual influence of the innermost and outermost planet are not included in $\Ha_{res}$ as this is a higher order effect. Note also that by dropping the higher order terms the problem is reduced to a planar one. 
In order to maintain the canonical nature of the equations of motion, we introduce for each planet the modified Delaunay action-angle variables $(\LAMBDONA,\GAMMONA,\LAMBDA,\GAMMA)$ (omitting the subscripts 1,2,3 for simplicity), which are given in terms of the orbital elements by
\begin{alignat}{2}\label{eq:ModifDelaunayVariables}
\LAMBDONA         &=\mu\sqrt{\GravC (M_*+m) a}\simeq m\sqrt{\GravC M_* a},   &&\quad \LAMBDA = \M + \VARPI, \nonumber\\
\GAMMONA          &=\LAMBDONA(1-\sqrt{1-e^2})\simeq\LAMBDONA e^2/2,       &&\quad \GAMMA = -\VARPI,
\end{alignat}
where $\mu = M_* m/(M_*+m)$ is the reduced mass, 
and $\M = \E - e\sin\E$ is the mean anomaly ($\E$ being the eccentric anomaly).
In these variables, the Keplerian part $\Ha_{kepl}$ of the Hamiltonian \eqref{eq:FullHamiltonian} takes the form
\begin{equation}\label{eq:KeplerianPartInModifDelaunayVariables}
\Ha_{kepl} = -\sum_{i=1}^3 \frac{\GravC^2 (M_* + m_i)^2 \mu_i^3}{2\LAMBDONA_i^2} \simeq -\sum_{i=1}^3 \frac{m_i^3}{2}\left(\frac{\GravC M_*}{\LAMBDONA_i}\right)^2,
\end{equation}
while the resonant Hamiltonian writes
\begin{equation}\label{eq:ResonantHamiltonianInModifDelaunayVariables}
\begin{split}
\Ha_{res}		&= 		-\frac{\GravC^2 M_* m_1 m_2^3}{\LAMBDONA_2^2} \\
				&\quad\quad \times\left(f_{res}^{(1,in)}\sqrt{\frac{2\GAMMONA_1}{\LAMBDONA_1}} 
				\cos\big(k^{in}\LAMBDA_2-(k^{in}-1)\LAMBDA_1 + \GAMMA_1\big)\right.\\
				&\quad\quad\quad \left.+ f_{res}^{(2,in)}\sqrt{\frac{2\GAMMONA_2}{\LAMBDONA_2}} 
				\cos\big(k^{in}\LAMBDA_2-(k^{in}-1)\LAMBDA_1 + \GAMMA_2\big)\right) + \\
			&\quad 	-\frac{\GravC^2 M_* m_2 m_3^3}{\LAMBDONA_3^2} \\
				&\quad\quad \times\left(f_{res}^{(1,out)}\sqrt{\frac{2\GAMMONA_2}{\LAMBDONA_2}} 
				\cos\big(k^{out}\LAMBDA_3-(k^{out}-1)\LAMBDA_2 + \GAMMA_2\big)\right.\\
			&\quad\quad\quad \left.+ f_{res}^{(2,out)}\sqrt{\frac{2\GAMMONA_3}{\LAMBDONA_3}} 
				\cos\big(k^{out}\LAMBDA_3-(k^{out}-1)\LAMBDA_2 + \GAMMA_3\big)\right);
\end{split}
\end{equation}
note that in going from \eqref{eq:ResonantHamiltonianInOrbitalElements} to \eqref{eq:ResonantHamiltonianInModifDelaunayVariables} we have made use of the approximation $e\simeq\sqrt{2\GAMMONA/\LAMBDONA}$, which holds at first order in $e$.

This Hamiltonian is clearly not integrable. However, one can perform a series of changes of variables that allow us to reduce by two the number of degrees of freedom. 
The first canonical transformation is 
\begin{equation}\label{eq:ResHarmonicsCoV}
\begin{split}
\KAPPONA        &=\LAMBDONA_1 + \frac{k^{in}-1}{k^{in}}\LAMBDONA_2 + \frac{(k^{in}-1)(k^{out}-1)}{k^{in} k^{out}}\LAMBDONA_3,	\quad\quad \KAPPA = \LAMBDA_1, \\
\THETAONA^{(1)}	&=\frac{1}{k^{in}}\LAMBDONA_2 + \frac{k^{out}-1}{k^{in} k^{out}}\LAMBDONA_3, \quad\quad \THETA^{(1)}=k^{in}\LAMBDA_2-(k^{in}-1)\LAMBDA_1, \\
\THETAONA^{(2)}	&=\frac{1}{k^{out}}\LAMBDONA_3, \quad\quad \THETA^{(2)}=k^{out}\LAMBDA_3-(k^{out}-1)\LAMBDA_2; 
\end{split}
\end{equation}
it is straightforward to check using the Poisson bracket criterion \citep{2002mcma.book.....M} that it is indeed canonical.
Now, the new angle $\KAPPA$ does not appear explicitly in the Hamiltonian, which makes $\KAPPONA$ a constant of motion. 
The significance of $\KAPPONA$ has already been discussed for two planets (e.g.\ \citealt{2008MNRAS.387..747M}, \citealt{2013A&A...556A..28B}), and it has to do with the location of exact resonance. As we have already mentioned, neighbouring planets can still be in resonance while their semi-major axis ratios do not satisfy exactly the resonant condition $a_i/a_{i+1}=((k-1)/k)^{2/3}$, therefore the observed $a_{i,obs}$ do not alone reveal how close the planets are to resonance, nor do they represent the nominal $\bar{a}_i$ that do satisfy it. However by calculating from $a_{i,obs}$ the value of the constant of motion $\KAPPONA$, and imposing in the formula
\begin{equation}
\begin{split}
\frac{\KAPPONA}{\LAMBDONA_3} &= \frac{\mu_1}{\mu_3}\sqrt{\frac{M_*+m_1}{M_*+m_3}\frac{a_1}{a_3}} + \frac{k^{in}-1}{k^{in}}\frac{\mu_2}{\mu_3}\sqrt{\frac{M_*+m_2}{M_*+m_3}\frac{a_2}{a_3}} + \\
&\quad\quad+\frac{(k^{in}-1)(k^{out}-1)}{k^{in} k^{out}},
\end{split}
\end{equation} 
the condition of exact resonance, $a_i/a_{i+1}=((k-1)/k)^{2/3}$, for all pairs $i=1,2$, we derive the nominal value of $\bar\LAMBDONA_3$.
From this, one easily obtains $\bar{a}_3$ from $\bar{a}_3 = (\bar\LAMBDONA_3/m_3)^2 / (\GravC M_*)$, and then recursively $\bar{a}_2=((k^{out}-1)/k^{out})^{2/3} \bar{a}_{3}$, and finally $\bar{a}_1=((k^{in}-1)/k^{in})^{2/3} \bar{a}_{2}$. 

It is worth briefly recalling here why even in resonance the planets' semi-major axes do not coincide exactly with their nominal values. As an example, for the inner planet, the condition for resonance is that the resonant angle $k^{in} \LAMBDA_2 -(k^{in}-1)\LAMBDA_1 - \VARPI_1$ is librating, $0\sim k^{in} \dot\LAMBDA_2 - (k^{in}-1)\dot\LAMBDA_1 - \dot\VARPI_1 = k^{in} n_2 - (k^{in}-1) n_1 -\dot\VARPI_1$, which together with the condition of exact nominal resonance $k^{in} n_2 - (k^{in}-1) n_1 = 0$ would imply $\dot\VARPI_1\sim0$; however from the Hamiltonian \eqref{eq:ResonantHamiltonianInModifDelaunayVariables} we have $\dot\VARPI_1=-\dot\GAMMA_1\propto\GAMMONA_1^{-1/2}\sim 1/e_1$ which grows as $e_1\searrow 0$, meaning that at low eccentricities $\dot\VARPI_1\nsim 0$, which in turn forces $k^{in} \dot\LAMBDA_2 - (k^{in}-1)\dot\LAMBDA_1 = k^{in} n_2 - (k^{in}-1) n_1 \nsim 0$ in order to maintain the libration of the resonant angle. The resonant equilibrium points will therefore correspond to semi-major axes $a_i$ which may well deviate farther and farther from $\bar a_i$ as $e_i$ approaches 0. 
We can however already greatly simplify the calculations given that we will only consider deviations of the semi-major axis ratios from the nominal ones of no more than 5\% (moreover, very small values of the eccentricities are observationally disfavoured for Kepler systems, \citealt{2017AJ....154....5H}). In this limit, we can expand the Keplerian part to second order in $\delta\LAMBDONA_i=\LAMBDONA_i-\bar\LAMBDONA_i$, where $\bar\LAMBDONA_i=\mu_1\sqrt{\GravC(M_*+m_i)\bar a_1}$ is the nominal resonant value of $\LAMBDONA_i$, and write
\begin{equation}
\begin{split}
\Ha_{kepl} &= -\sum_{i=1}^3 \frac{\GravC^2 (M_* + m_i)^2 \mu_i^3}{2} \\
&\quad\quad \times\left(\frac{1}{\bar\LAMBDONA_i^2} - 2\frac{1}{\bar\LAMBDONA_i^3}\delta\LAMBDONA_i + 3\frac{1}{\bar\LAMBDONA_i^4}\delta\LAMBDONA_i^2 + \mathcal O(\delta\LAMBDONA_i^3)  \right),
\end{split}
\end{equation}
which, inserting the definition of $\delta\LAMBDONA_i$ and dropping the unimportant constant term and the higher order terms, reduces to:
\begin{equation}\label{eq:ExpandedKeplerianHamiltonian}
\Ha_{kepl} = \sum_{i=1}^3\left(4\bar n_i \LAMBDONA_i - \frac{3}{2} \bar h_1 \LAMBDONA_i^2\right),
\end{equation}
where $\bar n_i = \sqrt{\GravC(M_*+m_i)/\bar a^3}$ is the nominal mean motion and $\bar h_i = \bar n_i/\bar\LAMBDONA_i = 1/(\mu_i \bar a_i^2)$ can be interpreted as the inverse of the moment of inertia of a circular orbit around the star.
As we will see below, for the purposes of our study, considering the expanded Keplerian Hamiltonian up to order $\mathcal O(\delta\LAMBDONA^2)$ does not introduce any significant inaccuracy in our calculations.
Concerning the resonant Hamiltonian \eqref{eq:ResonantHamiltonianInModifDelaunayVariables}, we can evaluate it at the nominal values $\LAMBDONA=\bar\LAMBDONA$ as it is already of order $\mathcal O(e)$.

Finally, one last canonical change of variable is made:
\begin{alignat}{2}\label{eq:FinalCoV}
\OMEGONA           &=\THETAONA^{(1)} + \THETAONA^{(2)} - (\GAMMONA_1 + \GAMMONA_2 + \GAMMONA_3),      &&\quad \THETAP=\theta^{(1)}, \nonumber \\
\PSIONA_1^{(1)}    &=\GAMMONA_1 + \GAMMONA_2 + \GAMMONA_3 - \THETAONA^{(2)},  &&\quad \PSI_1^{(1)} = \THETA^{(1)}+\GAMMA_1, \nonumber\\
\PSIONA_1^{(2)}    &=\THETAONA^{(2)},  &&\quad \PSI_1^{(2)} = \THETA^{(2)}+\GAMMA_2, \nonumber\\
\PSIONA_2^{(1)}    &=-\GAMMONA_2-\GAMMONA_3+\THETAONA^{(2)},  &&\quad \DELTAGAMMA^{(1)} = \GAMMA_1-\GAMMA_2, \nonumber\\
\PSIONA_2^{(2)}    &=-\GAMMONA_3,  &&\quad \DELTAGAMMA^{(2)} = \GAMMA_2-\GAMMA_3.
\end{alignat}
Again, we see that the new angle $\THETAP$ does not appear in the Hamiltonian, making $\OMEGONA$ another constant of motion of the system (note that here $\OMEGONA$ does not denote the longitude of the node which does not appear in our model, since the problem is planar). We are therefore left with a four-degree-of-freedom Hamiltonian which depends parametrically on the constants of motion $\KAPPONA$, $\OMEGONA$. We already mentioned the meaning of $\KAPPONA$; for $\OMEGONA$, one can easily show that $\KAPPONA+\OMEGONA = (\LAMBDONA_1-\GAMMONA_1)+(\LAMBDONA_2-\GAMMONA_2)+(\LAMBDONA_3-\GAMMONA_3) \equiv \ANGMOM$, the total angular momentum of the system, which is to be expected knowing that it is a conserved quantity.

\subsection{Resonant equilibrium points}\label{subsec:ResonantEquilibriumPoints}
Let us briefly summarise our work so far. We have obtained a 4-degrees-of-freedom Hamiltonian $\bar\Ha$ which is a function of the actions $\big(\PSIONA_1^{(1)}, \PSIONA_1^{(2)},\PSIONA_2^{(1)},\PSIONA_2^{(2)}\big)$ and the angles $\big(\PSI_1^{(1)}, \PSI_1^{(2)},\DELTAGAMMA^{(1)},\DELTAGAMMA^{(2)}\big)$, and depends parametrically on the values of $\KAPPONA$ and $\OMEGONA$ (which are linked to the orbital elements as expressed in \eqref{eq:ModifDelaunayVariables} and \eqref{eq:FinalCoV}); the Hamiltonian in these variables reads
\begin{equation}\label{eq:ReducedHamiltonianWithExpandedKeplerianPart}
\begin{split}
\bar\Ha &= \Ha_{kepl} + \Ha_{res},\\
\Ha_{kepl} &= \Ha_{kepl}\left(\PSIONA_1^{(1)},\PSIONA_1^{(2)};\KAPPONA,\OMEGONA\right),\\
\Ha_{res} &= \Ha_{res}\left(\PSIONA_1^{(1)}, \PSIONA_1^{(2)},\PSIONA_2^{(1)},\PSIONA_2^{(2)},\PSI_1^{(1)}, \PSI_1^{(2)},\DELTAGAMMA^{(1)},\DELTAGAMMA^{(2)};\KAPPONA,\OMEGONA\right) ,
\end{split}
\end{equation}
where the explicit dependence of each term can be obtained by direct substitution.
We now consider the stable equilibria of this system. We look for equilibrium points of this Hamiltonian by simultaneously solving the set of equations
\begin{equation}\label{eq:EquationsForResonantEquilibria}
\begin{split}
\frac{\partial\bar\Ha}{\partial\PSIONA_1^{(1)}}&=0,~\frac{\partial\bar\Ha}{\partial\PSIONA_1^{(2)}}=0,~\frac{\partial\bar\Ha}{\partial\PSIONA_2^{(1)}}=0,~\frac{\partial\bar\Ha}{\partial\PSIONA_2^{(2)}}=0,\\
\frac{\partial\bar\Ha}{\partial\PSI_1^{(1)}}&=0,~\frac{\partial\bar\Ha}{\partial\PSI_1^{(2)}}=0,~\frac{\partial\bar\Ha}{\DELTAGAMMA^{(1)}}=0,~\frac{\partial\bar\Ha}{\DELTAGAMMA^{(2)}}=0.
\end{split}
\end{equation}
Note that by the functional form of the Hamiltonian, any combination of values in $\{0,\pi\}$ for the angles immediately satisfies the last line. These are known as the \emph{symmetric equilibria}. Asymmetric equilibria are possible (e.g.\ \citealt{2006MNRAS.365.1160B}), but they do not play a role at the low eccentricities at which we are limiting ourselves here. 

Plugging in specific values for the angles in $\{0,\pi\}$ reduces the problem of solving the four equations that appear in the first line to find the stable equilibria of the system.
Note that although the Hamiltonian depends on both $\OMEGONA$ and $\KAPPONA$, the latter assumes a natural value for any specific problem at hand (that is, any values of $m_1$, $m_2$, $m_3$ and of $k^{in}$, $k^{out}$) by rescaling the units so that e.g.\ $\bar a_1=1$. To trace out the loci of the resonant equilibria, we then simply change the value of $\OMEGONA$ (which corresponds to changing the angular momentum $\ANGMOM$, at constant $\KAPPONA$) and solve equations \eqref{eq:EquationsForResonantEquilibria} to find $\big(\PSIONA_{1,eq}^{(1)}, \PSIONA_{1,eq}^{(2)},\PSIONA_{2,eq}^{(1)},\PSIONA_{2,eq}^{(2)}\big)$, which are then translated into orbital elements working backwards through the canonical transformations.

\begin{figure}
\centering
\includegraphics[width=0.4\textwidth 
]{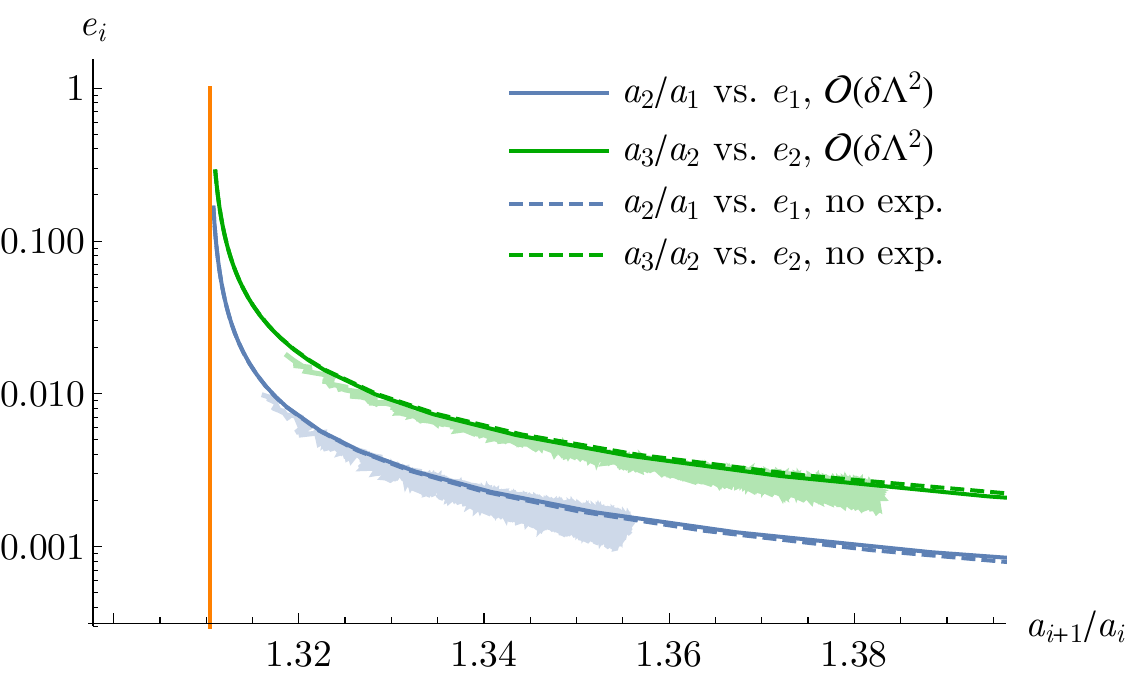}
\caption{
Equilibrium curves showing the loci of the stable resonant equilibrium points calculated as explained in the text, in the case of a 3:2 -- 3:2 mean motion resonance chain, with $m_1=m_2=m_3=10^{-4} M_*$. The full curves are calculated using the expanded Keplerian Hamiltonian \eqref{eq:ExpandedKeplerianHamiltonian}, while the dashed curves are calculated using the unexpanded Keplerian Hamiltonian \eqref{eq:KeplerianPartInModifDelaunayVariables}, showing very little difference down to very small eccentricities and for reasonable values of the nearly exactly resonant semi-major axis ratio. The location of the nominal resonant semi-major axis ratio $(3/2)^{2/3}$ is shown by a vertical orange line. We also superimpose the numerically computed evolution of a 3-planet system deep in the 3:2 mean motion resonance (for both pairs) and undergoing dissipative evolution depicted with transparent lines: the system follows the locations of the equilibrium points, which are close to the curves calculated analytically. 
}
\label{fig:3Planets_3-2_EqualMass_EquilibriumCurves} 
\end{figure}

We show in Figure \ref{fig:3Planets_3-2_EqualMass_EquilibriumCurves} one example of equilibrium curves for three equal-mass planets down to eccentricities of order $10^{-3}$, where we also show that the expanded Keplerian Hamiltonian provides an accurate description of the system. These curves are then matched against the result of full $N$-body numerical simulations of a system with the same physical parameters which starts deep in resonance and evolves dissipatively so to slowly follow the resonant equilibrium points (transparent lines).\footnote{
Note that even away from nominal resonance all four resonant angles can continue to librate when the system is sufficiently close to the resonant equilibrium point, unlike what has been erroneously stated in Section 4 of \cite{2013AJ....145....1B}. 
}
These $N$-body integrations with the addition of dissipative effects will be detailed below in Section \ref{sec:RealSystemsStudy.subsec.Numerical}.

\subsection{Resonant repulsion for three-planets systems}\label{subsec:ResonantRepulsionForThreePlanets}
The equilibrium curves in the $a_{i+1}/a_i$ vs $e_i$ plane show that the resonant repulsion during energy dissipation is expected also for three-planets systems.
For systems of two planets, it is well known that for first order resonances the resonant equilibria always reside wide of the exact resonant ratio of the semi-major axes. That is, because the resonant condition requires $k\dot\LAMBDA_2-(k-1)\dot\LAMBDA_1-\dot\VARPI_{1,2}\simeq0$ and since the perihelion precession is always retrograde, $\dot\VARPI_{1,2}<0$, one necessarily has $k\dot\LAMBDA_2-(k-1)\dot\LAMBDA_1<0$ i.e.\ $a_2/a_1>(k/(k-1))^{2/3}$. 
More concretely, at low enough eccentricities and at semi-major axis ratios close to the nominal ones, one finds directly using the resonant Hamiltonian expanded to first order in $e$ that $\dot\VARPI_1=\alpha f_{res}^{(1)}n_1 (m_2/M_*)e_1^{-1}$, $\dot\VARPI_2 = -f_{res}^{(2)} n_2 (m_1/M_*)e_2^{-1}$ with $f_{res}^{(1)}<0$, $f_{res}^{(2)}>0$: this means that the lower are the eccentricities, the wider are the equilibria from the exact commensurability. At higher eccentricities the secular terms, of $\mathcal O(e^2)$, become more important, and they contribute a positive contribution (that is constant in $e$) in $\dot\VARPI$; however, one still finds $\dot\VARPI_{1,2}<0$ at higher eccentricities as well (e.g.\ \citealt{2018CeMDA.130...54P}).

For systems with three planets, since we used a first order expansion in $e$ in the analytical model and therefore we are not including the mutual interaction between the inner planet and the outer planet, the perihelion precession will still be retrograde and it will remain true that for each pair of planets the resonant equilibria lie wide of the exact nominal resonance, and that, in the limit of small enough eccentricities, the separations grow with diminishing eccentricities. This is indeed what we see in Figure \ref{fig:3Planets_3-2_EqualMass_EquilibriumCurves}, where the analytically computed resonant equilibria agree very well with our numerical simulations.\footnote{
At higher eccentricities the main term which might shift the equilibria in $a_2/a_1$, $a_3/a_2$ to the left of exact resonance is the second order (secular) term which describes the interaction between the inner planet and outer planet; however, we checked that adding this term to the Hamiltonian, even at high eccentricities and for a very massive outer planet the picture does not change.} 

Consider now a resonant 3-planet system that is close to some resonant equilibrium point and is subjected to (tidal) dissipation. Assuming that the dissipative evolution is slow compared to that of the resonant variables, which has a characteristic timescale given by the libration period at vanishing amplitude of libration, the system will remain bound to the equilibrium curves. Since the eccentricities are damped by the dissipation, we conclude that the semi-major axes are expected to diverge.
We also conclude that systems of three planets that are close to a given first-order mean motion resonance but for which one or both pairs is narrow of the resonance can only be explained by a resonant configuration if the amplitude of libration around the resonant equilibrium point is large, and temporarily takes the planets to period ratios that are lower than the exact resonant period ratios.

We finally note here a property of these curves that will be used later.
The Hamiltonian \eqref{eq:ReducedHamiltonianWithExpandedKeplerianPart} can be rescaled by a parameter which encapsulates all of the information regarding how the dynamics scales with mass ratios and physical sizes of the orbits. This is analogous to the rescaling found e.g.\ in \cite{2013A&A...556A..28B} for the 2-planets case, and only works when using an expanded Hamiltonian and for semi-major axes close to the nominal resonant ones, which are our working assumptions anyway.
Therefore, after rescaling all planetary masses by a certain factor $\tilde m$, the corresponding loci of the resonant equilibria are also simply rescaled, and can be immediately calculated. More specifically, one can easily see that for given semi-major axis ratios, the values of the eccentricities that correspond to the resonant equilibrium point are just rescaled by $\tilde m$, since the eccentricities and the planetary masses appear as a product in the perturbing Hamiltonian. This can be easily understood using the previous formula $\dot\VARPI\propto (m_{pl}/M_*)e^{-1}$, and noticing that fixing the semi-major axis ratio ultimately fixes $\dot\VARPI$ by the resonance condition; therefore, rescaling the planetary masses, at fixed $\dot{\VARPI_1}=\dot{\VARPI_2}=\dot{\VARPI_3}$ (i.e.\ at constant semi-major axis ratios), the eccentricities are simply rescaled by the same factor. 
This implies that for a given equilibrium configuration of the semi-major axis ratio $a_{2,eq}/a_{1,eq}$, the corresponding equilibrium of the ratio $a_{3,eq}/a_{2,eq}$ will be independent of $\tilde m$, that is independent of the absolute value of the planetary masses. Only the \emph{ratios} $m_1/m_2$ and $m_2/m_3$ are significant, meaning that if one changes one of these ratios, the equilibrium in $a_{3,eq}/a_{2,eq}$ corresponding to the same $a_{2,eq}/a_{1,eq}$ will have changed (see the Appendix \ref{appendix:ReducedHamiltonian} for an explicit presentation of this rescaling procedure).

\section{A scenario for dissipative evolution of 3-planet systems}\label{sec:TestingScenario}
In this section, we select near-resonant systems of 3 planets from the NASA Exoplanet Archive catalogue, and discuss whether or not their observed orbital configuration is compatible with the dynamical evolution driven by the following physical processes.
As we already mentioned, planets are expected to dynamically interact with the protoplanetary disk in which they formed. This has two effects: a damping of the eccentricities (and inclinations), and an exchange in angular momentum between the planets and the disk which causes the planets' semi-major axes to change (planetary migration). Planets captured in mean motion resonance are usually attributed to inward migration (that is, the planet looses orbital angular momentum to the disk), and we will consider this case in this paper, but these results are general to any form of convergent evolution. In our implementation, when the inner planet reaches the inner edge of the disk, migration is stopped, and the second incoming planet will cross a commensurability with the first; finally, the third planet will cross a commensurability with the second. We note that the timescale over which the eccentricity is damped is usually of order $\sim100$ times shorter than that over which the semi major axis changes. Planets are therefore expected to approach these commensurabilities with vanishingly low eccentricity. As we can see from Figure \ref{fig:3Planets_3-2_EqualMass_EquilibriumCurves}, however, this configuration with semi-major axis ratio wide of resonance and low eccentricity is very close to the resonant equilibrium points. 
The planets keep approaching each other and their eccentricities keep increasing due to the curvature of the locus of the equilibrium points in the $(a_{i+1}/a_i,e_i)$ diagram, until an equilibrium is reached when the damping of $e$ balances such resonant eccentricity pumping.
This is why planets are expected to be (close to) a resonant equilibrium point in the first place, and a chain of resonances can be formed.
The disk is then slowly depleted and the planets maintain their configuration. 

Following the depletion of the gas, dissipation is introduced which removes orbital energy at constant angular momentum; this is done here implementing tidal dissipation but again the method is general.   
During this phase of dissipative evolution, the planets will follow again the equilibrium curves of the resonant Hamiltonian for changing $\OMEGONA$, this time decreasing their eccentricities and hence increasing the semi major axis ratios $a_{i+1}/a_i$ for each planet-planet pair. Note that $\OMEGONA$ changes because $\KAPPONA$ changes (since do the semi-major axes as a consequence of the dissipation of energy) and $\ANGMOM$ has to stay constant for this kind of dissipation.

\subsection{Choice of systems}\label{subsec:ChoiceOfSystems}
Recall that the only orbital parameters that are well constrained by transit data are the orbital periods, which allow us to obtain the semi-major axis ratios even without knowing the mass of the star. The orbital periods listed in the NASA Exoplanet Archive catalogue are, for the cases considered below, obtained by fase folding the observed signal. Since we will be considering short-period planets this is equivalent to obtaining the proper value of the periods, so that we can directly compare the corresponding observed semi-major axis ratios with the ones coming from our averaged model of resonance.\footnote{
We should remark however that even the periods are not known with arbitrary precision, meaning that there might be small discrepancies in the values that are used in different works. In this paper, we will use the ones listed in the NASA Exoplanet Archive catalogue without considering error bars. This is enough for the scope of our analysis.} 
The masses of the planets could be obtained starting from the estimates for the planetary radii, and making use of a scaling relation such as the one found in \cite{2013ApJ...772...74W}, $m_{pl}=3 m_\oplus (R_{pl}/R_\oplus)$, where $m_{pl}$, $R_{pl}$ are the mass and radius of the exoplanet, and $R_\oplus$, $m_\oplus$ those of the Earth. However this is only a statistical law and the uncertainties on the mean densities usually preclude accurate estimates for the masses, which are indeed not yet known. We will therefore use the masses as free parameters of our study.
In fact, as we already saw, the only significant quantities for our study are the mass \emph{ratios} between the planet; this follows from the discussion at the end of Section \ref{subsec:ResonantRepulsionForThreePlanets}. In practice, we will choose a total planetary mass $m_{tot}=m_1+m_2+m_3$ by using for each system the mean planetary radius $(R_1+R_2+R_3)/3$, the aforementioned scaling relationship from \cite{2013ApJ...772...74W} to obtain an average planetary mass $m_{pl,avg}$, and setting $m_{tot}=3 m_{pl,avg}$. Again, this is simply a choice that we are forced to make in order to run $N$-body simulations, but it does not in any way change our result, which is therefore not sensitive to the uncertainties on the radii (or to the lack of their knowledge, as will be the case for YZ Ceti). Note however that since individual Kepler systems seem to show a homogeneity in planetary radii and masses \citep{2018AJ....155...48W,2017ApJ...849L..33M}, this choice likely constitutes a good approximation to the real architecture of these systems.

Given a pair of neighbouring planets with periods $P_1$ and $P_2$ respectively, which are close to a given first order resonance, $P_1/P_2\simeq (k-1)/k$, one can define (see for example \citealt{2012ApJ...761..122L,2014ApJ...787...80H})
\begin{equation}
\Delta_{k/(k-1)}:=\frac{k-1}{k}\frac{P_2}{P_1}-1,
\end{equation}
called the \emph{normalised distance from (exact) resonance}. Note that when $\Delta>0$ the planets reside \emph{wide} of the $k:k-1$ resonance, while when $\Delta<0$ the planets are \emph{narrow} of the resonance. 

\begin{figure}[!t]
\centering
\includegraphics[width=0.4\textwidth 
]{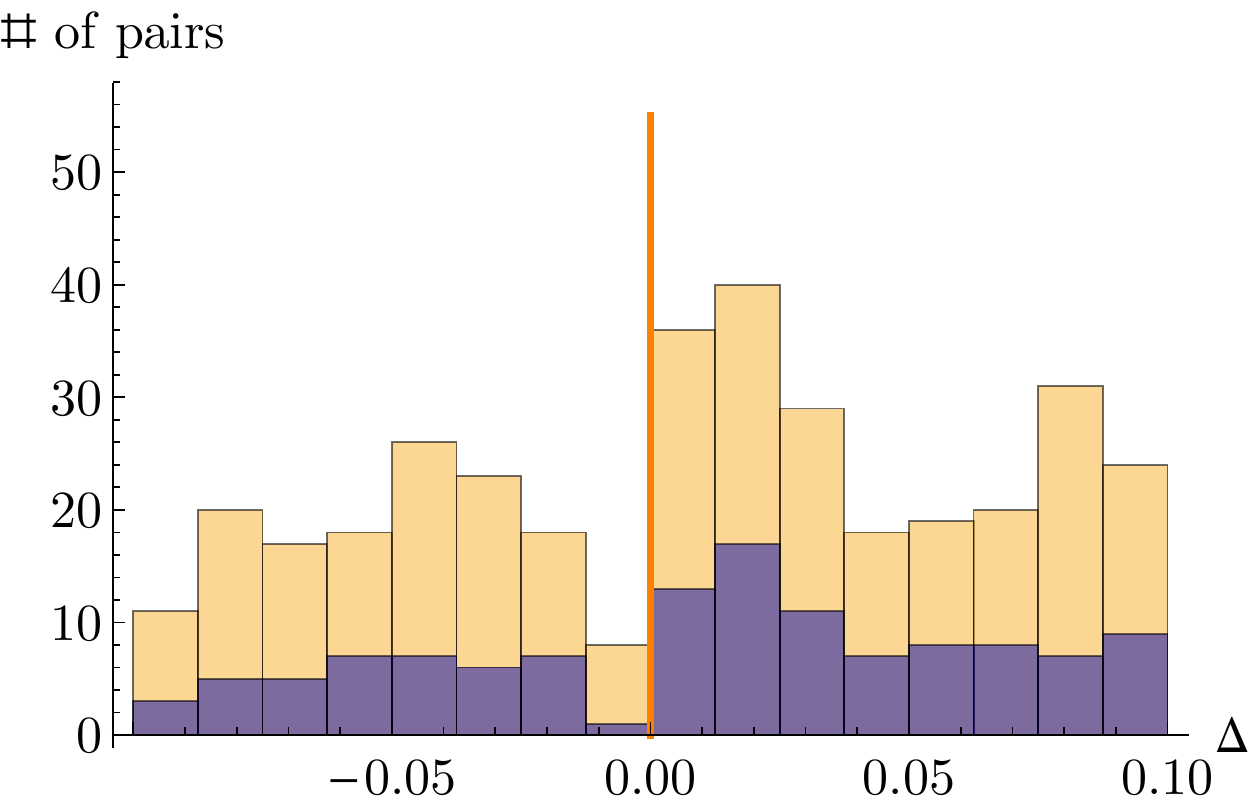}
\caption{Distribution of the (signed) normalised distance from first order mean motion resonances $\Delta=\frac{k-1}{k}\frac{P_2}{P_1}-1$ with $k=2,3$ in all exoplanetary systems (yellow) and for 3-planet systems (purple). We note a clear peak to the right of the value $\Delta=0$ (corresponding to the exact nominal resonance, indicated by a vertical orange line), which is most prominent for $0<\Delta\lesssim 0.05$. Out of the 358 pairs plotted here, there are 123 total pairs in this configuration. For the 3-planets systems only, 48 pairs have $0<\Delta\lesssim 0.05$, out of the 121 shown in the purple histogram.}
\label{fig:DeltDistr}
\end{figure}

We will be selecting planetary systems of three planets with both pairs close to some first order mean motion resonance such that $|\Delta|\lesssim0.05$ holds for both pairs, with $k=2,3$ as they appear to be the most common resonances. 
We recall that the normalised width of a resonance is of order $\sim (m/M_*)^{2/3}$ (\citealt{2013ApJ...774..129D,2015MNRAS.451.2589B}), 
where the average planetary mass for Kepler systems is of order $m\sim 3\times 10^{-5}M_*$, and moreover most planets in each system appear to be quite homogeneous in mass \citep{2018AJ....155...48W,2017ApJ...849L..33M}. This gives a typical resonance width of order $\Delta\sim 10^{-3}$ in normalised units, meaning that in selecting systems with $0<\Delta \lesssim 0.05$ we are generously including planetary pairs with separation 50 times larger than the typical resonant width. Moreover, the available data shows that for systems close to mean motion resonance, the distribution of $\Delta$ favours values between $0$ and $\lesssim 0.05$ (\cite{2017AJ....154....5H}, see also Figure \ref{fig:DeltDistr}). Additionally, we will require that $\Delta>0$, which is justified by our results in the previous section. 

\begin{figure}[!ht]
\centering
\includegraphics[width=0.5\textwidth 
]{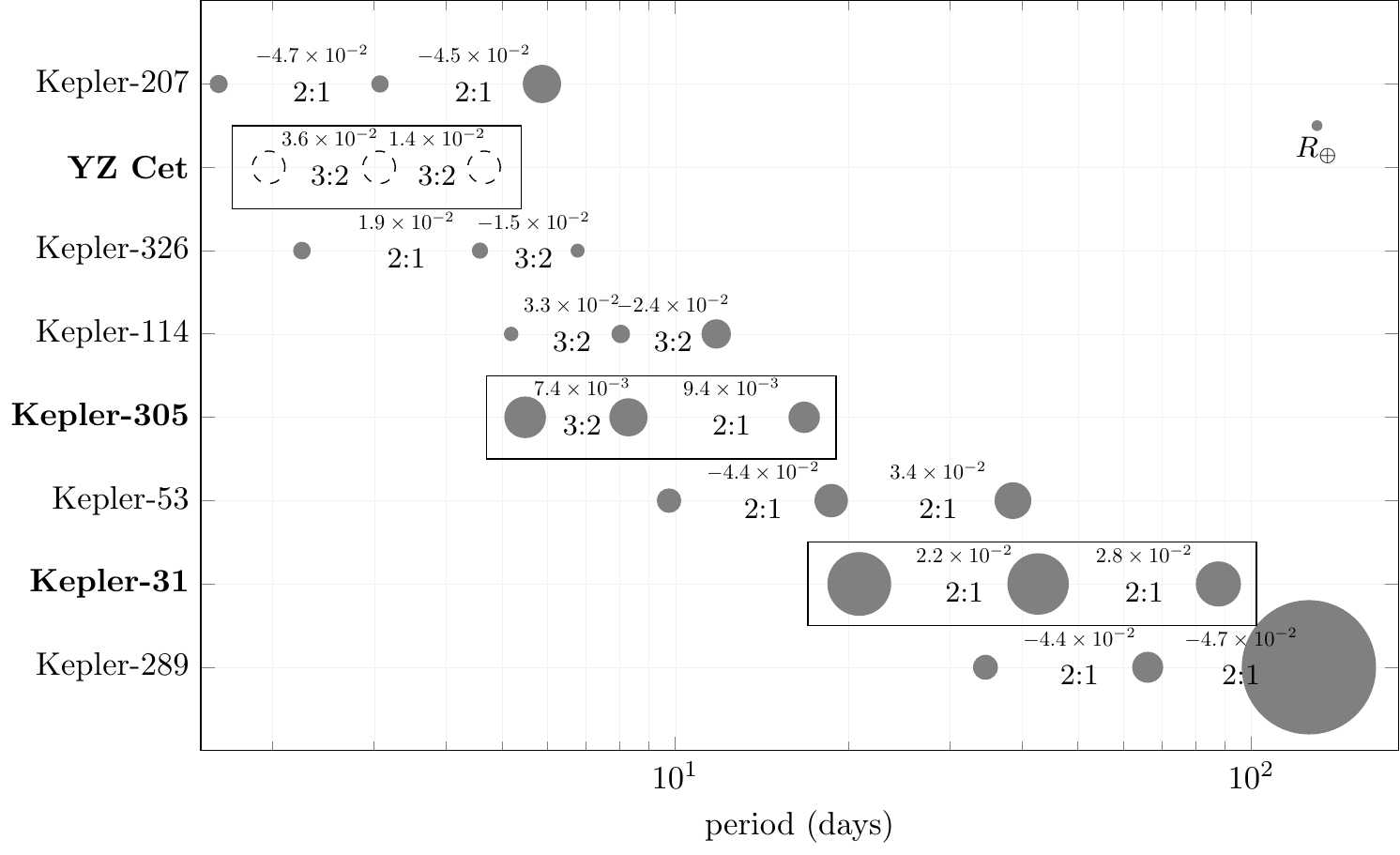}
\caption{Orrery of the 3-planets systems sufficiently close to first order mean motion resonances $k$:$k-1$ with $k=2,3$, with a normalised distance to resonance $|\Delta|<0.05$ for \emph{both} pairs. For each system, we place a circle in correspondence of the period of each planet, and indicate between pairs of planets the first order mean motion resonance in which they are envisioned to reside (below) and the normalised distance to that resonance $\Delta$ (above); the sign of $\Delta$ indicates if the pair of planets are narrow ($\Delta<0$) or wide ($\Delta>0$) of the resonance. For our analysis, we will only consider systems for which $\Delta>0$ for both pairs (the systems enclosed by a box). The size of the circle is an indication of the estimated radius of the planet (the small dot in the top right corner demonstrates the size of Earth). For YZ Ceti this information is not available, but this does not pose a problem for our study.}
\label{fig:Orrery}
\end{figure}

Of all 3-planet systems, only 8 satisfy $|\Delta|<0.05$ for both pairs, that is, appear to be close to a multi-resonant chain (they are 
Kepler-31, 
Kepler-53, 
Kepler-114, 
Kepler-207, 
Kepler-289, 
Kepler-305, 
Kepler-326, 
YZ Ceti). The architecture of these systems is shown in Figure \ref{fig:Orrery}; of these, only 3 satisfy $\Delta>0$ for both pairs. These are 
Kepler-31, 
Kepler-305 and
YZ Ceti.
For these systems, we consider whether or not their observed orbital configuration can be consistent with the scenario envisioned above. 

\subsection{Analytical Maps}\label{sec:RealSystemsStudy.subsec.Analytical}

With our analytical model of resonance at hand, we can construct analytical maps of resonant equilibrium points for different resonant chains and different planetary mass ratios. For the purposes of the current study, we proceed as follows. For an arbitrary system, we assume to have observations for the orbital period ratios and obtain the values of $k^{(in)}$ and $k^{(out)}$. We then pick both mass ratios $m_2/m_1$ and $m_3/m_2$ and construct the equilibrium curves as explained in Section \ref{subsec:ResonantEquilibriumPoints} (in practice, we work with the aforementioned Hamiltonian rescaled by the common planetary mass factor $\tilde m$, see the Appendix \ref{appendix:ReducedHamiltonian}). Then, we find the resonant equilibrium point (i.e.\ the value of $\ANGMOM$) that corresponds to a value of $a_3/a_1$ equal to the observed semi-major axis ratio, and therefore put $\left.(a_3/a_1)\right|_{eq}\equiv\left.(a_3/a_1)\right|_{obs}$. The determined value of $\ANGMOM$ fixes the eccentricities of all three planets, since they are all linked by the equilibrium curves.

Recall that we only have \emph{one} free parameter to select the chosen equilibrium configuration: the angular momentum; however, we have \emph{two} observables that we want to match, which are both semi-major axis ratios $a_2/a_1$ and $a_3/a_2$. This is unlike the case of only two resonant planets, where one has \emph{one} free parameter (again the angular momentum $\ANGMOM$) and only \emph{one} observable (the single $a_2/a_1$ ratio): in this case, it would \emph{always} be possible to find a suiting value of $\ANGMOM$ which gives a resonant equilibrium configuration such that the semi-major axis ratio is equal to the observed one (provided that the latter is wide of the nominal value $(k/(k-1))^{2/3}$). 
In the case of three planets, choosing $\ANGMOM$ such that $\left.(a_3/a_1)\right|_{eq}\equiv\left.(a_3/a_1)\right|_{obs}$ automatically fixes the equilibrium values of both semi-major axis ratios $a_2/a_1$ and $a_3/a_2$. Considering for example the corresponding equilibrium  $\left.(a_3/a_2)\right|_{eq}$, we obtain the weighted difference
\begin{equation}\label{eq:WeightedDifferenceOfSmaRatio}
\bar\delta (a_3/a_2): = \frac{(\left.(a_3/a_2)\right|_{eq}-\left.(a_3/a_2)\right|_{obs})}{\left.(a_3/a_2)\right|_{obs}}
\end{equation}
between $\left.(a_3/a_2)\right|_{eq}$ and the observed value $\left.(a_3/a_2)\right|_{obs}$. The same can be done for $a_2/a_1$, which gives a similar (absolute) result, $|\bar\delta (a_2/a_1)|\simeq |\bar\delta (a_3/a_2)|$. Maintaining this procedure, we loop over different planetary mass ratios.\\

\begin{figure*}[!ht]
\centering
\begin{subfigure}[b]{0.45 \textwidth}
\centering
\includegraphics[scale=0.6
]{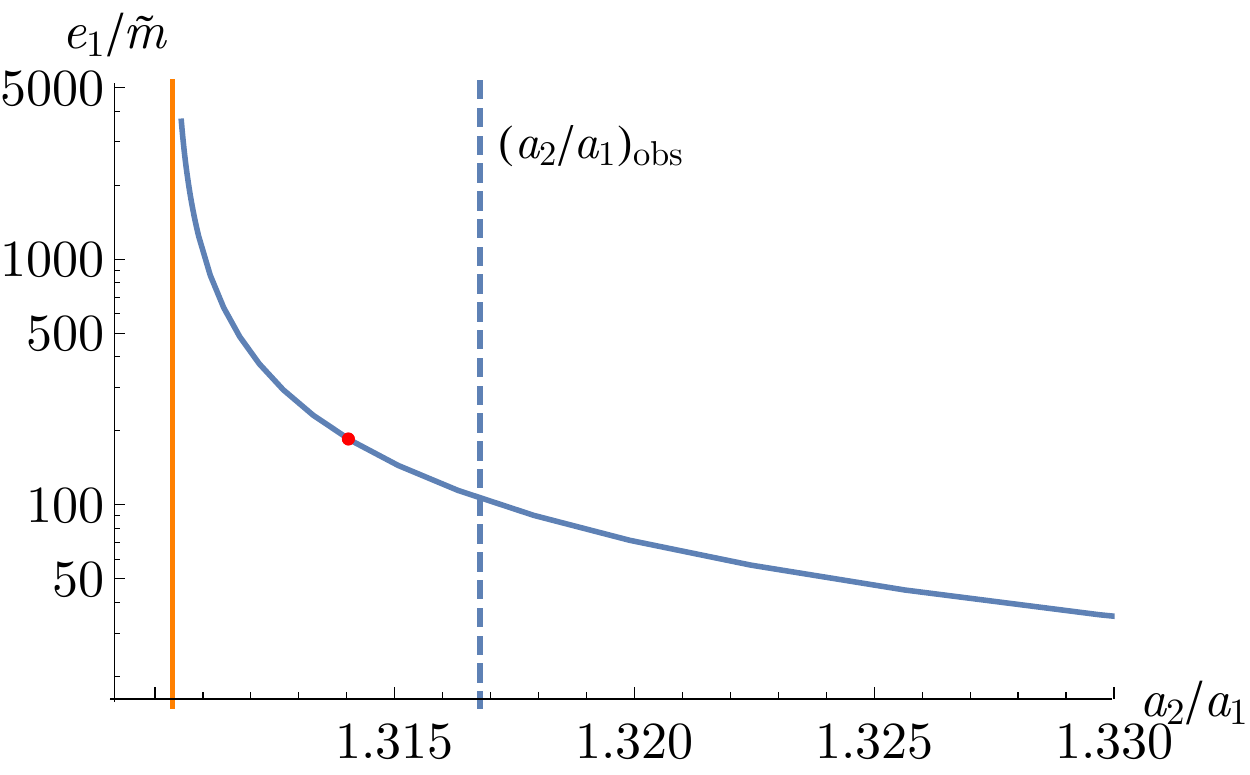}
\end{subfigure}
\hfill
\begin{subfigure}[b]{0.45 \textwidth}
\centering
\includegraphics[scale=0.6
]{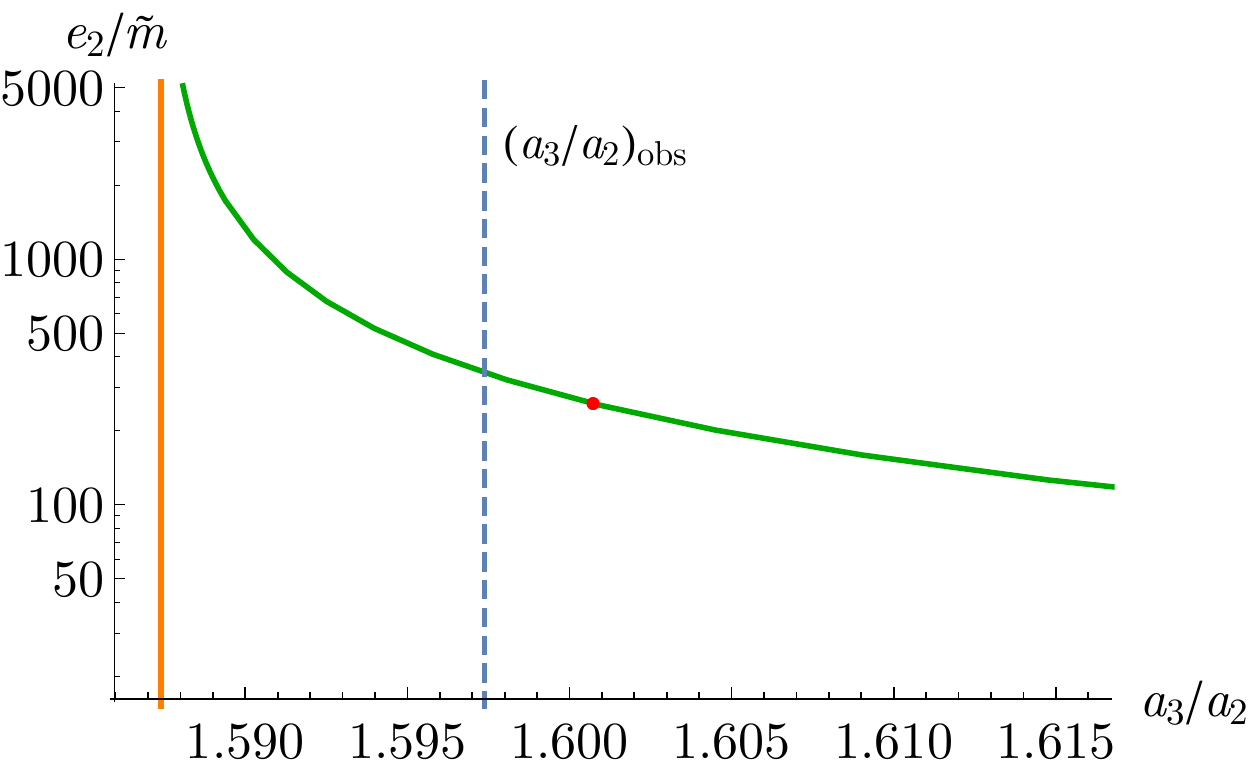}
\end{subfigure}



\caption%
{
Locations of the resonant equilibrium points in the $a_{i+1}/a_i$ vs.\ $e_i/\tilde m$ planes, $i=1,2$, for three equal-mass planets in a 3:2 -- 2:1 mean motion resonance chain, close to which Kepler-305 resides. Orange vertical lines show the exact nominal commensurability, while dashed vertical lines show the observed $a_2/a_1$ and $a_3/a_2$ in the case of Kepler-305. As explained in the text, we select \emph{one} equilibrium configuration (indicated by the red dot in \emph{both} panels) by requiring that $\left.(a_3/a_1)\right|_{eq}\equiv\left.(a_3/a_1)\right|_{obs}$, which automatically fixes \emph{all} orbital elements $a_2/a_1$, $a_3/a_2$, $e_1$, $e_2$ and $e_3$ along the equilibrium curves.
}
\label{fig:Kepler-305a2overa1Anda3overa2vse1_EqvsObs} 
\end{figure*}

\begin{figure*}[!ht]
\centering
\begin{subfigure}[b]{0.45 \textwidth}
\centering
\includegraphics[scale=.75]{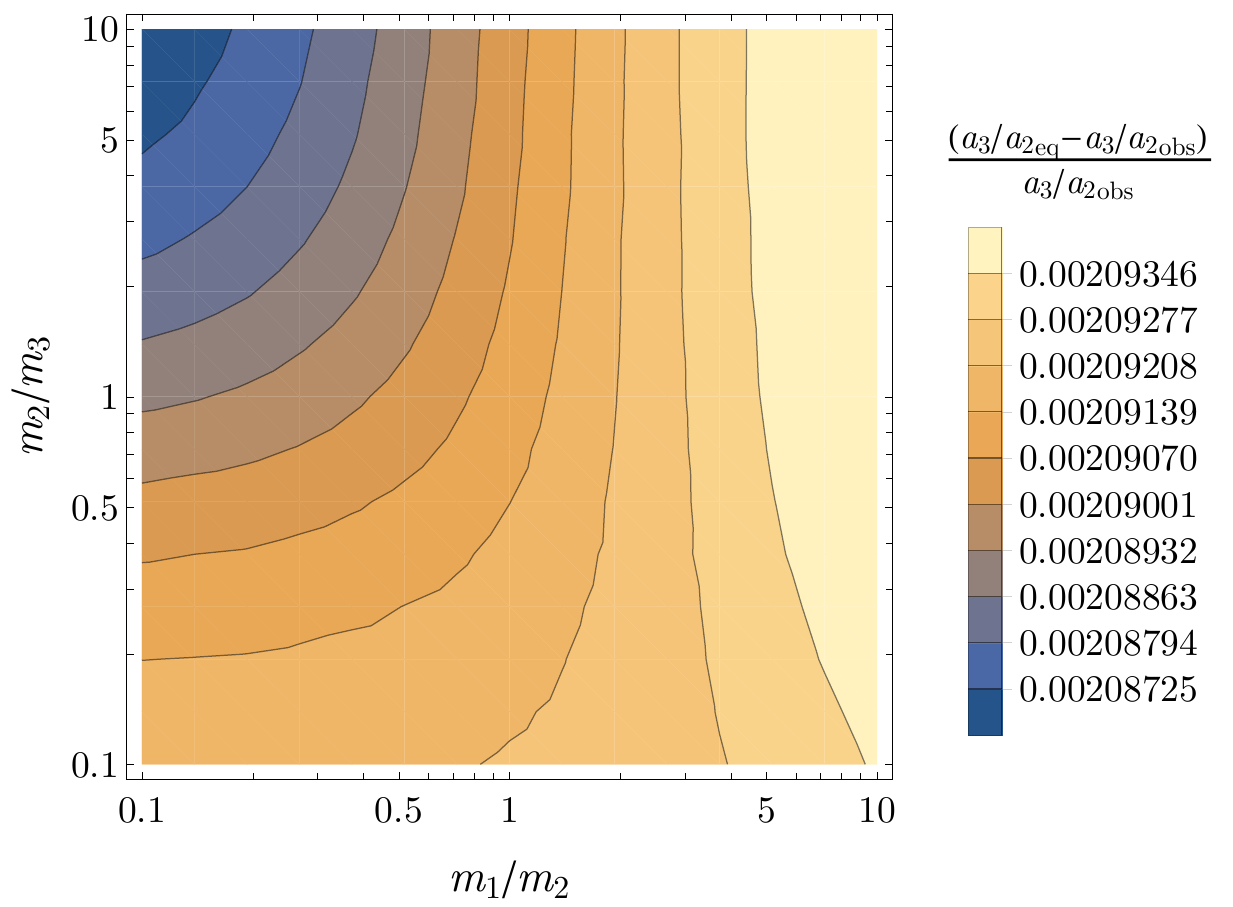}
\caption{}
\end{subfigure}
\hfill
\begin{subfigure}[b]{0.45 \textwidth}
\centering
\includegraphics[scale=.75]{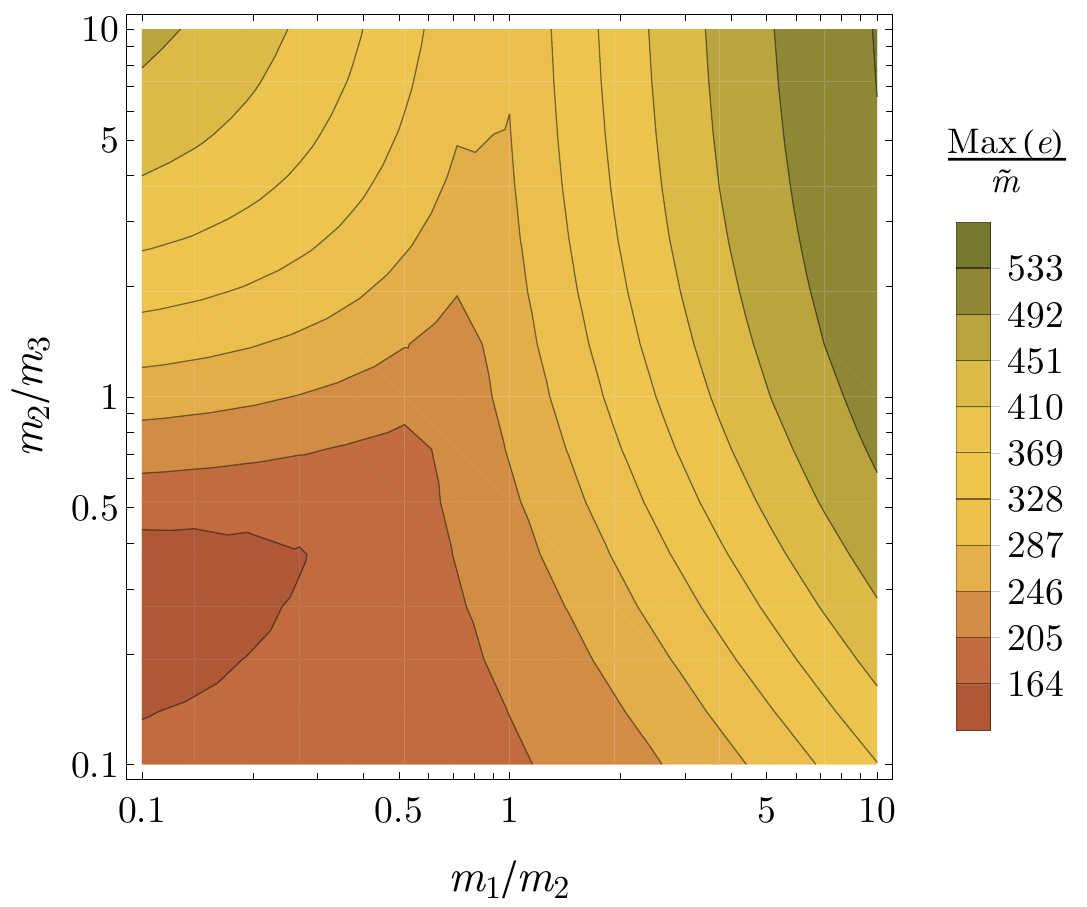}
\caption{}
\end{subfigure}

\vskip\baselineskip

\begin{subfigure}[b]{0.45 \textwidth}
\centering
\includegraphics[scale=.75]{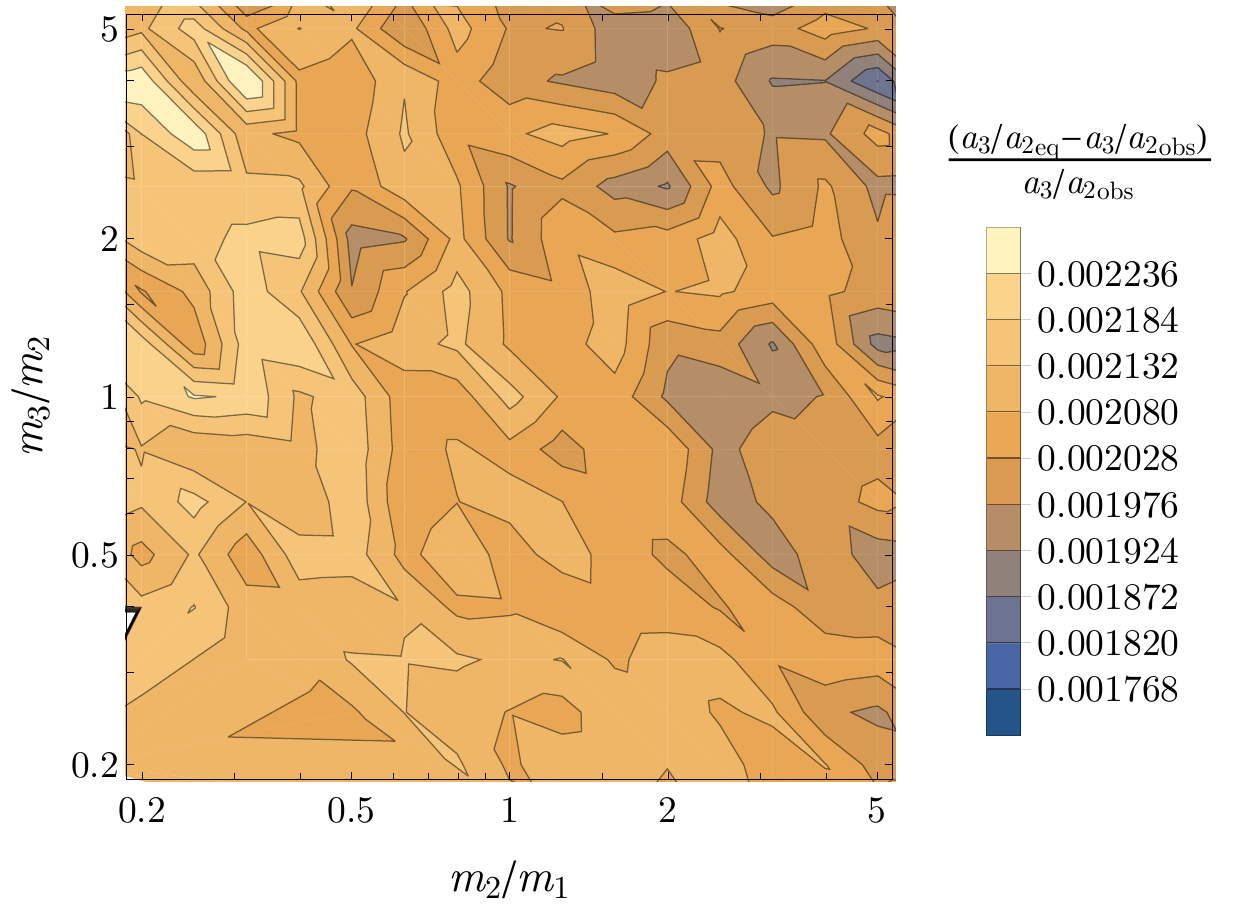}
\caption{}
\end{subfigure}
\hfill
\begin{subfigure}[b]{0.45 \textwidth}
\centering
\includegraphics[scale=.75]{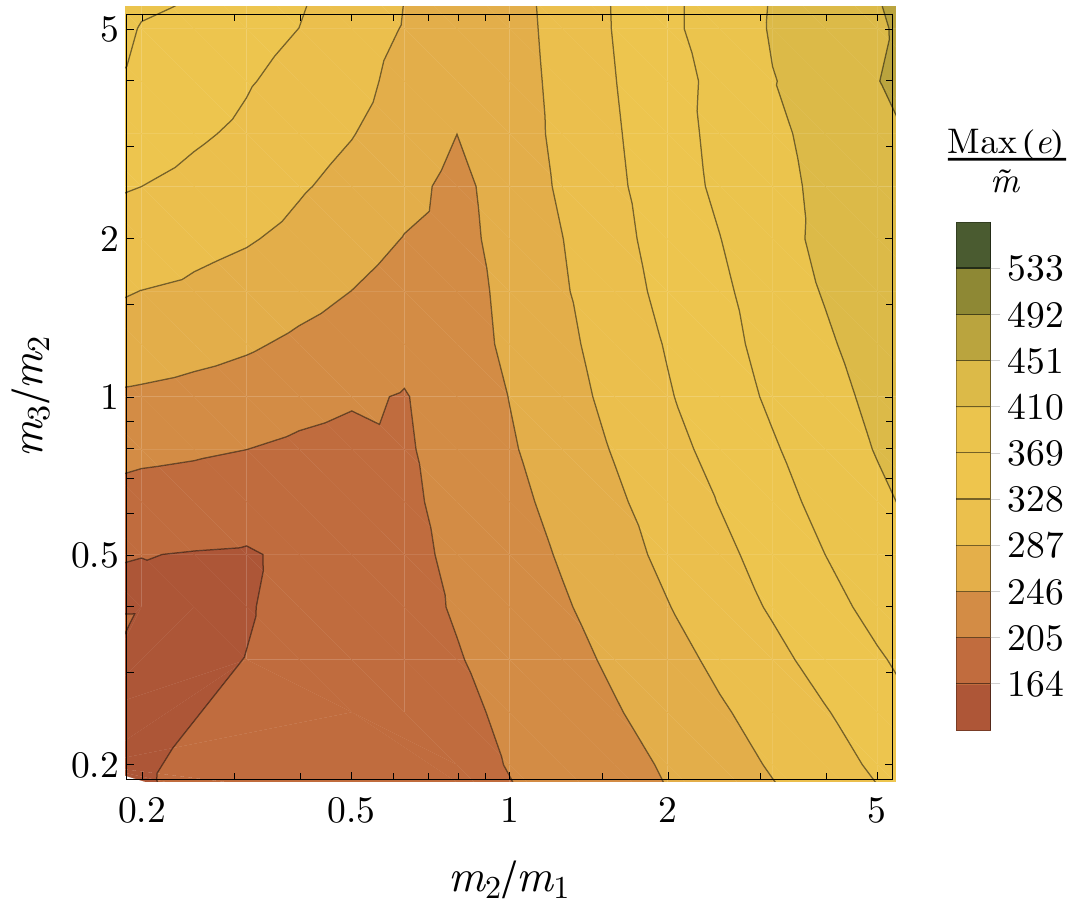}
\caption{}
\end{subfigure}

\caption%
{ 
{\it Top row}: Analytical maps constructed for Kepler-305 as explained in Section \ref{sec:RealSystemsStudy.subsec.Analytical}.
In panel (a) we plot the quantity $\bar\delta (a_3/a_2)$, which represents how close the system is now to some resonant equilibrium point, for different mass ratios $m_1/m_2$ and $m_2/m_3$ (each point in this plot is constructed by repeating the procedure described in Figure \ref{fig:Kepler-305a2overa1Anda3overa2vse1_EqvsObs}). We notice that $\bar\delta (a_3/a_2)$ changes very little with the mass ratios, and is of the order of $\sim0.002$. Comparing with Figure \ref{fig:StatDistrOfdeltaa3overa2}, we see that this can be the case by pure chance only in $\sim15\%$ of randomly generated systems close to the 3:2 -- 2:1 mean motion resonance chain. 
In panel (b) we show a map of the quantity $\max e_{eq}/\tilde m$ representing the equilibrium orbital configuration that is selected at each fixed value of $m_1/m_2$ and $m_2/m_3$ by imposing $\left.(a_3/a_1)\right|_{eq}\equiv\left.(a_3/a_1)\right|_{obs}$.
{\it Bottom row}: Numerical maps constructed for Kepler-305 as explained in Section \ref{sec:RealSystemsStudy.subsec.Numerical}.
We show numerical maps of $\bar\delta (a_3/a_2)$ in panel (c) and $\max e_{eq}/\tilde m$ in panel (d), analogous to the analytical plots above (over a slightly smaller range of mass ratios for simplicity). These are intended to validate the analytical maps, and show very good agreement between corresponding panels.
}
\label{fig:ResultsKepler305} 
\end{figure*}

\begin{figure*}[!ht]
\centering
\begin{subfigure}[b]{0.45 \textwidth}
\centering
\includegraphics[scale=.75]{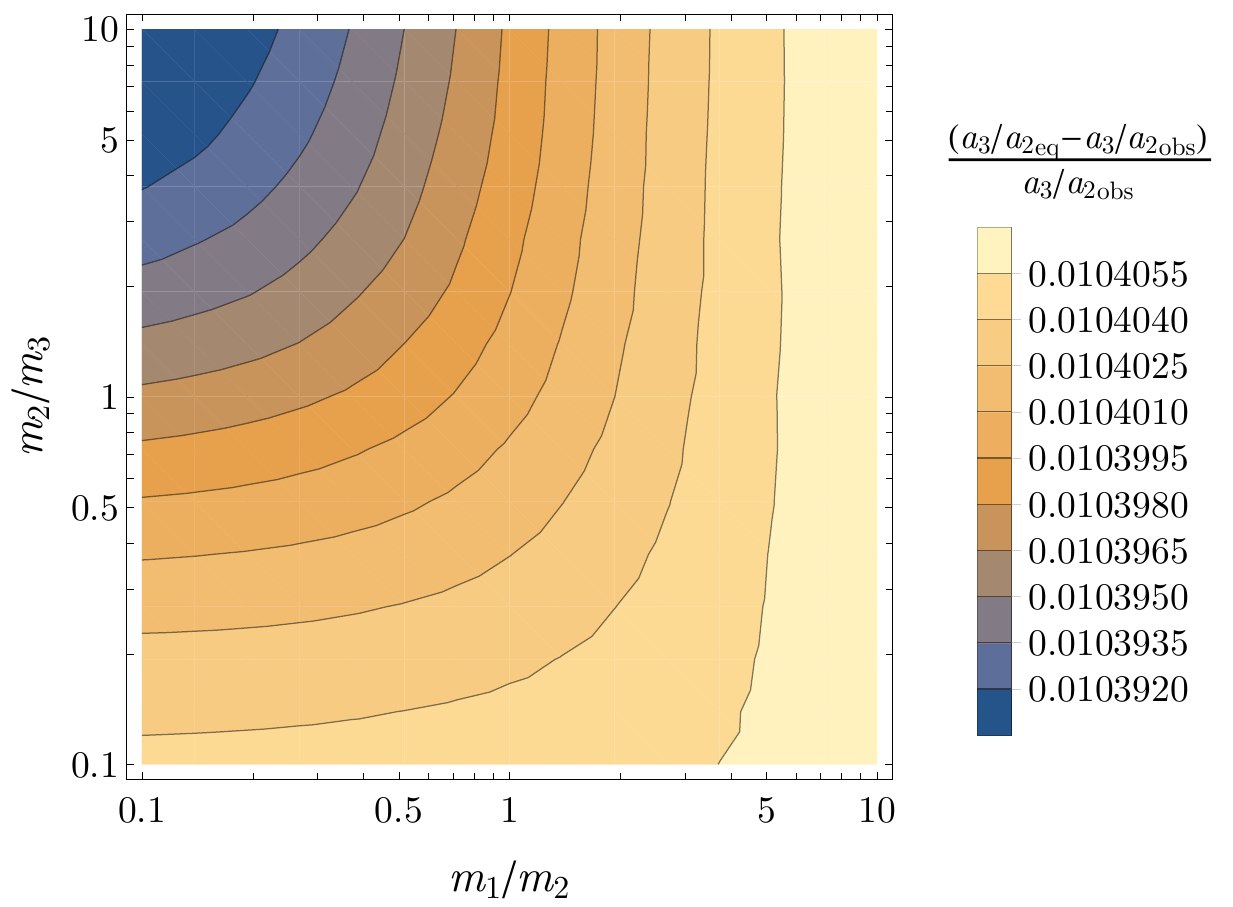}
\caption{}
\end{subfigure}
\hfill
\begin{subfigure}[b]{0.45 \textwidth}
\centering
\includegraphics[scale=.75]{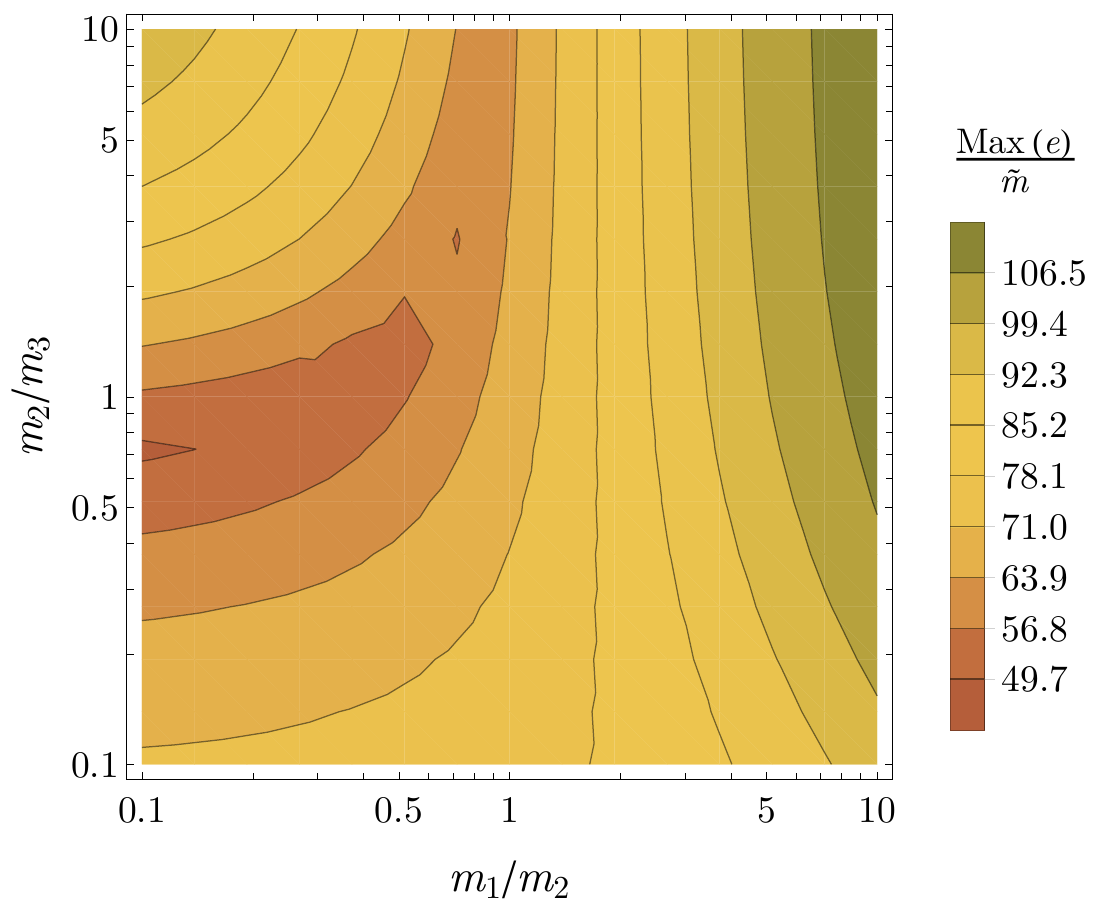}
\caption{}
\end{subfigure}

\vskip\baselineskip

\begin{subfigure}[b]{0.45 \textwidth}
\centering
\includegraphics[scale=.75]{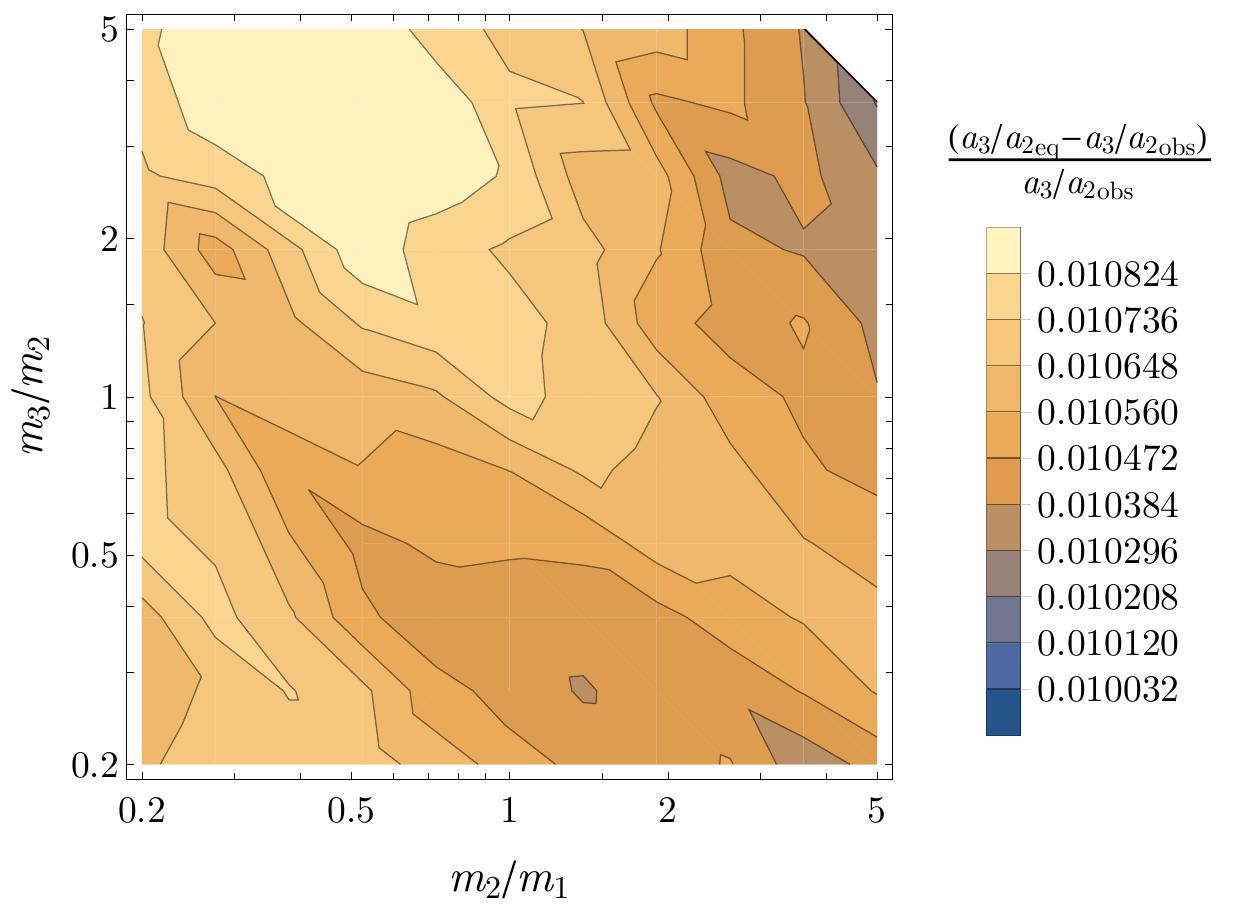}
\caption{}
\end{subfigure}
\hfill
\begin{subfigure}[b]{0.45 \textwidth}
\centering
\includegraphics[scale=.75]{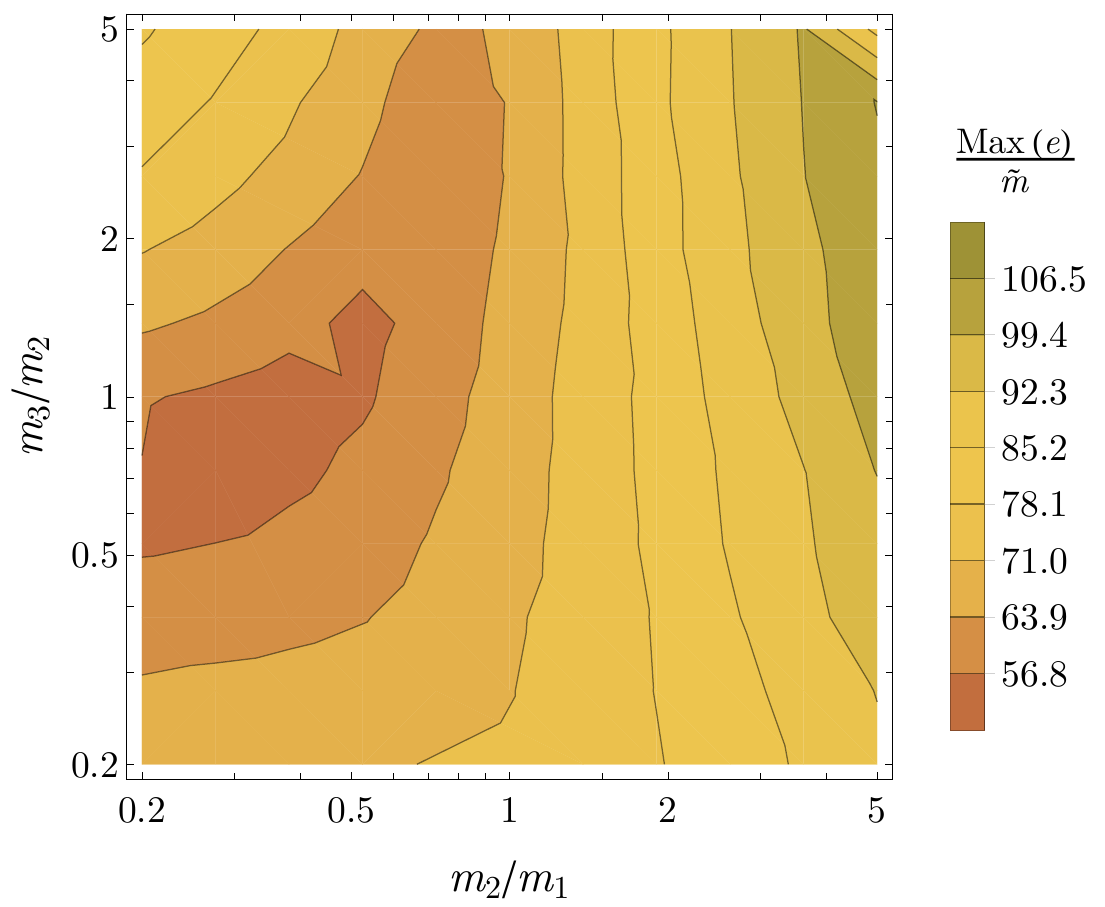}
\caption{}
\end{subfigure}

\caption%
{Same as in Figure \ref{fig:ResultsKepler305}, but for the system YZ Cet, residing close to a 3:2 -- 3:2 chain. The value of $\bar\delta (a_3/a_2)\sim0.01$ across all planetary mass ratios can be matched against the corresponding curve in Figure \ref{fig:StatDistrOfdeltaa3overa2}, where we find that $\sim80\%$ of randomly generated systems lie this close to the 3:2 -- 3:2 chain.
}
\label{fig:ResultsYZCet} 
\end{figure*}

\begin{figure*}[!ht]
\centering
\begin{subfigure}[b]{0.45 \textwidth}
\centering
\includegraphics[scale=0.6
]{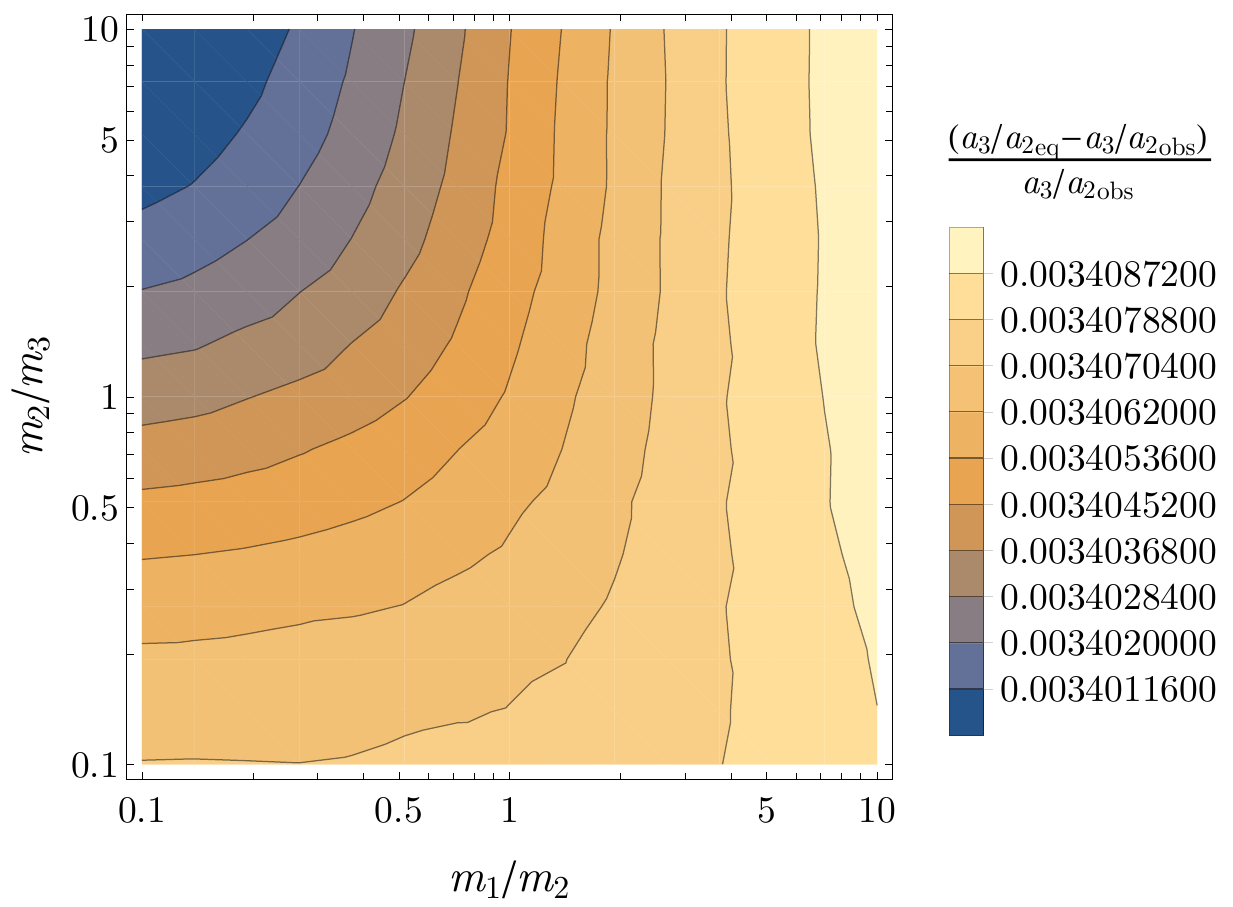}
\end{subfigure}
\hfill
\begin{subfigure}[b]{0.45 \textwidth}
\centering
\includegraphics[scale=0.6
]{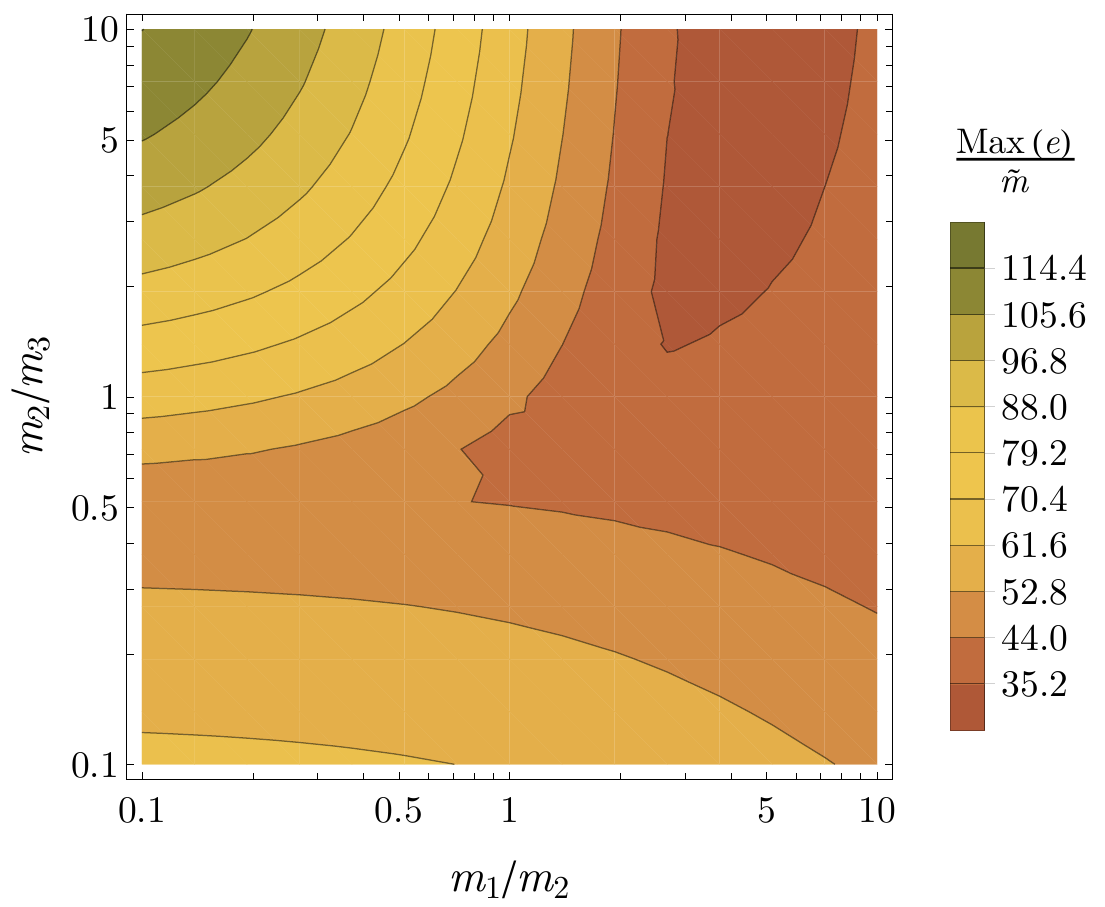}
\end{subfigure}



\caption%
{Same as in Figure \ref{fig:ResultsKepler305}, but for the system Kepler-31. Only analytical maps are shown in this case since in some simulations capture into resonance was unsuccessful due to overstability of the captured state for different planetary mass ratios. As explained in the text, this issue is model-dependent and is not within the scope of our analysis. Moreover, in the simulations where capture was successful, the results agree well with the analytical calculations, showing $\bar\delta (a_3/a_2)\sim0.003$. 
Comparing with Figure \ref{fig:StatDistrOfdeltaa3overa2} we see that there is a $\sim20\%$ probability that Kepler-31 lies this close to the 2:1 -- 2:1 chain by pure chance.
}
\label{fig:ResultsKepler31} 
\end{figure*}

We applied this procedure to the three systems selected above, starting with Kepler-305, which resides close to a 3:2 -- 2:1 mean motion resonance chain.
First of all, to better represent what these analytical maps intend to show, we draw in Figure \ref{fig:Kepler-305a2overa1Anda3overa2vse1_EqvsObs} equilibrium curves (equivalent to those shown in Figure \ref{fig:3Planets_3-2_EqualMass_EquilibriumCurves}) which describe the locations of the resonant equilibria for this resonant chain and for one specific choice of mass ratios $m_1/m_2=m_2/m_3=1$. The observed values of the semi-major axis ratios are indicated by dashed vertical lines; then, we indicate with a red dot the location of the specific equilibrium point that is selected when we impose $\left.(a_3/a_1)\right|_{eq}\equiv\left.(a_3/a_1)\right|_{obs}$; finally, we obtain $\bar\delta (a_3/a_2)$ as defined in \eqref{eq:WeightedDifferenceOfSmaRatio} (and similarly for $\bar\delta (a_2/a_1)$). As explained above, imposing $\left.(a_3/a_1)\right|_{eq}\equiv\left.(a_3/a_1)\right|_{obs}$ automatically fixes \emph{all} equilibrium eccentricities $e_{1,eq}$, $e_{2,eq}$, $e_{3,eq}$, and we can store their maximum $\max\{e_{1,eq},e_{2,eq},e_{3,eq}\}$ in to better describe the orbital configuration at the selected equilibrium point. We choose to consider the quantity $\max e_{eq}$ rescaled by a common planetary mass factor $\tilde m$ in order to obtain results that are independent of the planet-to-star mass ratio, since again the latter quantity does not effect equilibrium semi-major axis ratio configurations, and therefore does not affect $\bar\delta (a_3/a_2)$.
Panels (a) and (b) in Figure \ref{fig:ResultsKepler305} are the result of this procedure spanning different planetary mass ratios, showing maps of $\bar\delta (a_2/a_1)$ and $\max e/\tilde m$ using the observed semi-major axis ratios of Kepler-305 (the bottom panels (c) and (d) show the results of numerical simulations which will be detailed in Section \ref{sec:RealSystemsStudy.subsec.Numerical}, and are only intended to validate the analytical results). 
We show analogous results for the system YZ Cet, residing close to a 3:2 -- 3:2 chain, in Figure \ref{fig:ResultsYZCet}, and for For Kepler-31, a chain (close to the) 2:1 -- 2:1 mean motion resonances, in Figure \ref{fig:ResultsKepler31}.


\subsection{Numerical simulations}\label{sec:RealSystemsStudy.subsec.Numerical}
In order to check the validity of our analytical calculations, we turned to numerical simulations by performing the following study. 
We simulated the process of capture into a chain of first-order mean motion resonances by placing the planets relatively wide of the desired resonances, according to the specific values of $k^{in}$ and $k^{out}$ of each case, and simulating the effects of the protoplanetary disks by adding fictitious forces which mimic the interaction with the disk (\citealt{2006A&A...450..833C,2008A&A...482..677C}) to the $N$-body integrator {\codefont swift\_symba}. To ensure convergent migration for all planetary mass ratios, we stopped the migration of the inner planet by adding at the desired location a so-called planetary trap, which reproduces the effect of the inner edge of the disk and describes a disk cavity around a star (\citealt{2006ApJ...642..478M}, \citealt{2018CeMDA.130...54P}). As we mentioned above, the mass ratios were kept as free parameters. Since again we are interested in mass ratios of order unity, in our simulations we limited ourselves to $m_1/m_2$ and $m_2/m_3$ between $\sim 0.2$ and $\sim 5$, and repeated the same set of simulations. 

Recall that the timescale for planetary migration depends on the mass of the planet, meaning that changing the mass ratios will change the relative speeds at which each planet's semi-major axis decreases, which is practically inconvenient. Therefore, we used a fictitious $\tau_a$ which is kept equal for all planets and constant along the different simulations of resonant capture. This has the sole effect of making it easier to automate the simulations, and does not affect our results.
We need only to make sure that at the end of the disk-migration phase the semi-major axis ratios are smaller than the observed ones, since the subsequent evolution dominated by tidal dissipation will only cause the semi-major axis ratios to expand. Note that this is always possible, since one can obtain different final eccentricities at the captured state by changing the ratio of the eccentricity damping $\tau_e$ and the migration rate $\tau_a$, and thus obtain different corresponding equilibrium values of the semi-major axis ratios; the latter approach the exact commensurabilities as the eccentricities grow (Figure \ref{fig:3Planets_3-2_EqualMass_EquilibriumCurves}), and since for the systems that we are studying both pairs reside wide enough of the nominal resonance, the final eccentricities need not be too high, of order a few $10^{-2}$ for a typical planetary mass of order $10^{-5} M_*$.

After the desired resonant state is obtained, we slowly depleted the gas. Finally, we added the effects of tidal dissipation (following \citealt{2002ApJ...573..829M}), using arbitrary quality factors for the planets but large enough to ensure that the dissipative evolution be slow compared to the resonant evolution of the two planets' pairs. This allows us to perform efficient integrations without breaking the adiabatic approximation which keeps the system close to the resonant equilibrium points. 
Note that we have little to no information on the internal structures of exoplanets, so we would not be able to confidently assign realistic eccentricity damping timescales anyway. Moreover, as we have already mentioned, tides are only one example of dissipation (that is, loss of orbital energy $E$) at constant angular momentum $\ANGMOM$, so that these results are in fact generalisable to any dissipative process such that $\dot E < 0$ and $\dot\ANGMOM=0$.
Therefore, a resonant system undergoing any such process is expected to follow the loci of the resonant equilibria, and the divergence of the semi-major axes is obtained as a general result. 
We now explain how we obtain maps similar to those drawn in the previous section from these numerical simulations. 

Consider a choice of the mass ratios, and a simulation of the dissipative evolution of two pairs of resonant planets. The semi-major axis ratios $a_3/a_1$ and $a_3/_2$ will increase in time. When, for two consecutive outputs of the simulation, the ratio $a_3/a_1$ crosses the observed one $\left.(a_3/a_1)\right|_{obs}$, we store the corresponding value of $a_3/a_2$ from the simulation (an average of it at the two consecutive outputs). Since the system might be librating around the equilibrium points with some amplitude, and there are additional short period terms, this will happen many times for a single simulation, and we obtain a list of $a_3/a_2$ values. Then, we report the average of this list, and again compare this quantity with the observed $\left.(a_3/a_2)\right|_{obs}$ (since they are obtained from the mean period extracted from the data) by computing the relative difference as in \eqref{eq:WeightedDifferenceOfSmaRatio}. We then loop over different choices of planetary mass ratios and obtain a map that can be compared with the analytical maps of the previous section. A similar procedure can be applied to $a_2/a_1$ (which gives again similar values to that of $a_3/a_2$ as we mentioned in Section \ref{sec:RealSystemsStudy.subsec.Analytical}), as well as the quantity $\max e/\tilde m$.\\

This analysis has been performed for the three selected systems. For Kepler-305 and YZ Cet, we show the resulting plots on the bottom panels (c) and (d) of Figures \ref{fig:ResultsKepler305} and \ref{fig:ResultsYZCet} respectively, and notice very good agreement with the analytical results. The noise that is observed in the panels (c) relative to the quantity $\bar\delta (a_3/a_2)$ is due to the fact that the numerically simulated systems are undergoing fast oscillation while they cross the observed value $\left.(a_3/a_1)\right|_{obs}$, but the typical value of $\bar\delta (a_3/a_2)$ is similar to the one found analytically.

The case of Kepler-31, which resides close to a 2:1 -- 2:1 mean motion resonant chain, is a bit different, since the 2:1 resonance  capture might be only temporary if the librations around equilibrium are overstable \citep{2014AJ....147...32G}. 
This behaviour has been already investigated thoroughly in the case of two planets (\citealt{2015A&A...579A.128D}, \citealt{2015ApJ...810..119D}), however it has been shown to be dependent on the specific implementation of the disk-planet forces, and to disappear in some cases \citep{2018MNRAS.481.1538X}. In this work we do not intend to expand on these matters, since the formulas that mimic the planet-disk interactions represent only approximate implementations of the real forces that are felt by the planets from the disk, which themselves remain observationally unconstrained.
We therefore take a practical approach, and note that in the numerical simulations where resonant capture was successful (typically for $m_1/m_2, m_2/m_3 \gtrsim 1$) the numerical results agree very well with the analytical ones; moreover we still observe that the theoretical value of $\bar\delta (a_3/a_2)$ varies extremely little with $m_1/m_2$ and $m_2/m_3$ (Figure \ref{fig:ResultsKepler31}), so the latter simulations can be considered as enough support for the analytical calculations.

\section{Results}\label{sec:Results}

\subsection{Probabilistic measure of a resonant configuration in Kepler-305, YZ Cet and Kepler-31}
Using the maps of $\bar\delta (a_3/a_2)$ shown in Figures \ref{fig:ResultsKepler305}, \ref{fig:ResultsYZCet} and \ref{fig:ResultsKepler31} for the three selected systems Kepler-305, YZ Cet and Kepler-31, we draw the following conclusions.
First of all, one might expect that the quantity $\bar\delta (a_3/a_2)$ should change with the different choices of mass ratios, thus allowing one to make a prediction on their (so far unknown) values of $m_1/m_2$ and $m_2/m_3$ \emph{under the assumption} that these systems are indeed in resonance and evolving dissipatively. Follow-up monitoring of these systems could then produce new observations from which to obtain the real masses of the planets, and so validate or disprove the hypothesis.
However, in practice we find that these analytical maps show very little dependence on $m_1/m_2$ and $m_2/m_3$ spanning reasonable values. 
Note also that for all three systems $\bar\delta (a_3/a_2)$ is small, but never vanishing, which would represent an analytically computed equilibrium configuration such that $\bar\delta (a_3/a_2)=0$, that is, a resonant equilibrium point which satisfies $\left.(a_3/a_2)\right|_{eq}=\left.(a_3/a_2)\right|_{obs}$ and $\left.(a_2/a_1)\right|_{eq}=\left.(a_2/a_1)\right|_{obs}$. But even if this happened to be the case, the level curve $\bar\delta (a_3/a_2)=0$ would still span a broad range of mass ratios: given moreover the uncertainty in the observed period ratios of exoplanetary systems, this would make any determination of $m_2/m_1$ or $m_3/m_2$ using observed data, in general, inconclusive.

Secondly, we note that we do obtain in all three cases small values for $\bar\delta (a_3/a_2)$, meaning that these systems are indeed close to some equilibrium point of the Hamiltonian \eqref{eq:ReducedHamiltonianWithExpandedKeplerianPart} and therefore could potentially reside in a multi-resonant chain.
However, these values by themselves do not contain any meaningful information. 
The quantity $\bar\delta (a_3/a_2)$ should indeed be calibrated if we intend to use it as a measure of the probability that the actual system (with its real unknown eccentricities) is in resonance, which in turn would yield a measure of how consistent the orbital architecture of such a system is with the envisioned scenario described above. 
To this end, for various resonant chains we randomly generate period ratios of fictional systems such that $0<\Delta < 0.05$ for each pair, and extract the corresponding $\bar\delta (a_3/a_2)$ (calculated for the choice of mass ratios $m_1/m_2=m_2/m_3=1$ for simplicity, since as we saw above the $\bar\delta (a_3/a_2)$ value depends extremely weakly on the mass ratios). From the cumulative distribution of $|\bar\delta (a_3/a_2)|$ that arises from this procedure we can obtain the probability that any given system has a given (small) $\bar\delta (a_3/a_2)$ purely by chance.

\begin{figure}[!ht]
\centering
\includegraphics[width=0.4\textwidth 
]{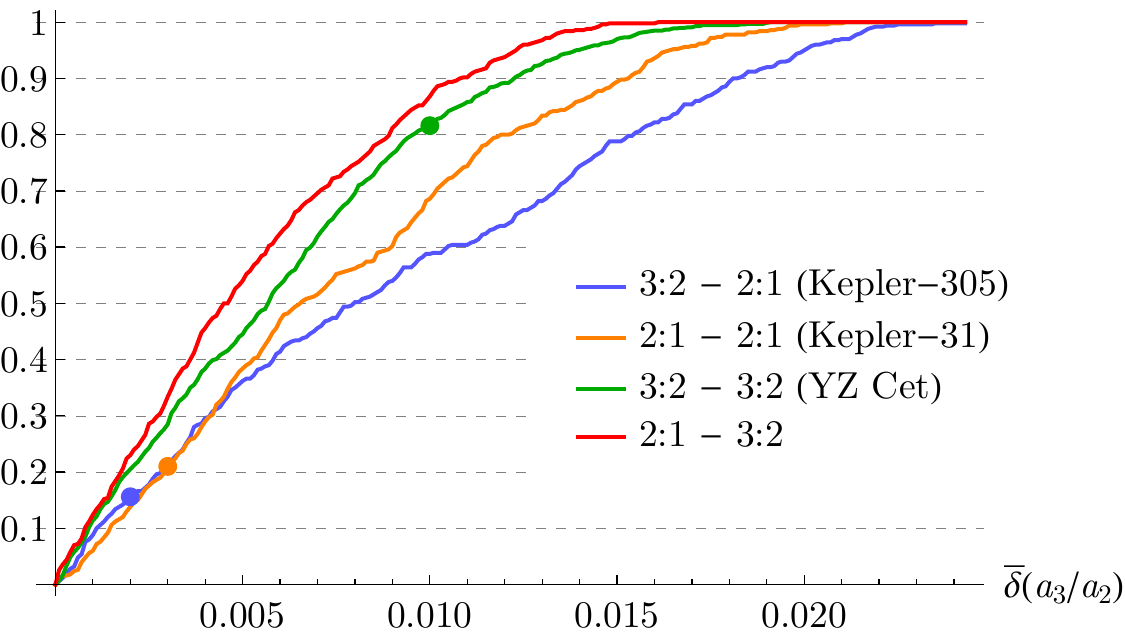}
\caption%
{
Cumulative distribution functions for $|\bar\delta (a_3/a_2)|$ for randomly generated systems close to chains of any possible combinations of the 2:1 and 3:2 mean motion resonances. We indicate the chains that represent selected systems from Figure \ref{fig:Orrery}; for each of them, a point indicates the observed $\bar\delta (a_3/a_2)$. From this, we obtain on the vertical axis the probability that these systems could have this value of $\bar\delta (a_3/a_2)$ by pure chance.
}
\label{fig:StatDistrOfdeltaa3overa2}
\end{figure}

Since we are interested mainly in the 2:1 and 3:2 mean motion resonances, we show in Figure \ref{fig:StatDistrOfdeltaa3overa2} these cumulative distributions for systems close to any possible combinations of these resonances.
The results show that the proximity of YZ Cet to the 3:2 -- 3:2 resonance is not statistically significant, since in $\sim80\%$ of randomly generate systems close to the 3:2 -- 3:2 chain we obtain an equivalent or smaller value of $\bar\delta (a_3/a_2)$. Instead, Kepler-305 and Kepler-31 are likely to be in resonance at the $1\sigma$ level (i.e.\ the probability that their value of $\bar\delta (a_3/a_2)$ is smaller than the determined value by chance is less than $32\%$): for Kepler-305 there is a $\sim15\%$ chance that this particular system lies this close to resonance by chance, while for Kepler-31 the probability is $\sim20\%$.

We should remark that these specific values for the probabilities that each system is this close to exact resonance by chance are calibrated by the choice $0<\Delta<0.05$ for both pairs of planets, which is used in generating the fictional systems. This value is however not arbitrary. For, it must be consistent with the choice made in Section \ref{subsec:ChoiceOfSystems}, which produced \emph{only} these three systems with \emph{both} the inner and outer planet pair this close to first order mean motion resonance: there, the choice $|\Delta|<\max\Delta=0.05$ was dictated by the observation of the location of the peak wide of nominal resonance (\cite{2017AJ....154....5H} and our Figure \ref{fig:DeltDistr}), so restricting the interval in $\Delta$ values with smaller $\max\Delta$ might have resulted in excluding potential systems. 
On the other hand, increasing $\max\Delta$ in the generation of the fictional systems would have the only effect to decrease the calculated probabilities. Therefore, we conclude that our choice of $\max\Delta=0.05$ is not arbitrary, and gives a reasonable upper bound to the probabilities that each system finds itself so close to exact resonance by pure chance.

For completeness, we report the observed variation of the three-body Laplace angle $\dot{\varphi_L}$ in these systems, since its libration can be in principle a sufficient condition to conclude that they are indeed resonant.
For Kepler-305 we checked that the Laplace angle $\varphi_L=2 \LAMBDA_1 - 4 \LAMBDA_2 + 2 \LAMBDA_3$ satisfies $\dot\varphi_L\simeq 0.5\degree\days^{-1}$ given the observed transits periods; for Kepler-31, $\varphi_L= \LAMBDA_1 - 3 \LAMBDA_2 + 2 \LAMBDA_3$ satisfies $\dot\varphi_L\simeq 0.1\degree\days^{-1}$; finally for YZ Cet, $\varphi_L= 2\LAMBDA_1 - 5 \LAMBDA_2 + 3 \LAMBDA_3$ satisfies $\dot\varphi_L\simeq 9.4\degree\days^{-1}$. 
As we argued in the Introduction, in case of libration of the resonant angles around the equilibrium point (and hence libration of the Laplace angle) the average of the $(a,e)$ oscillation falls on the equilibrium point, while in case of circulation, the average falls off the equilibrium point curve. Consequently, the circulation of the Laplace angle implies that the libration amplitude is larger than the distance of the equilibrium point from the axis $e=0$, and that $\bar\delta (a_3/a_2)$ cannot be zero. However, this does not mean that the system did not reach that point via divergent migration: being the $(a,e)$ equilibrium so close to to $e=0$, even a minute perturbation can induce circulation of the Laplace angle.
Hence the libration of the Laplace angle is a sufficient but not necessary condition to conclude that a system's dynamical history has been shaped by resonant capture and subsequent resonant repulsion driven by dissipation.

\subsection{The 5:4 -- 4:3 resonant chain on Kepler-60 and other near-resonant systems with $k>3$}
While in this work we have concentrated on the 2:1 and 3:2 mean motion commensurabilities, it is worthwhile to point out that more compact resonant chains are possible, and Kepler-60 represents a notable example. 
This system hosts three planets with mean observed periods of $\simeq5.49\days$, $\simeq8.29\days$ and $\simeq16.74\days$ respectively. Their masses have been constrained via TTV to be $\simeq4 M_\oplus$ for all planets \citep{2016ApJ...820...39J}. The mean motions of the planets satisfy $4\dot\LAMBDA_1-8\dot\LAMBDA_2+4\dot\LAMBDA_3\simeq -0.02 \degree\days^{-1}$, hinting at a resonant configuration. Indeed, \cite{2016MNRAS.455L.104G} found that the TTV signal for these planets is consistent with a true three-body Laplace-like resonance as well as a chain of 5:4 -- 4:3 two-body mean motion resonances.
Using the system's parameters we can find a resonant equilibrium configuration as in Section \ref{sec:RealSystemsStudy.subsec.Analytical}, by imposing $a_3/a_1$ to be equal to the observed $\left.(a_3/a_1)\right|_{obs}$. This gives $\bar\delta (a_3/a_2)$ of order $4\times 10^{-5}$. Using an analogous argument to that of Figure \ref{fig:StatDistrOfdeltaa3overa2}, we find that there is only a 0.25\% probability that Kepler-60 lies this close to a 5:4 -- 4:3 resonant chain by pure chance. The eccentricities that we find at the selected resonant equilibrium point are of order $e_1\simeq0.02$, $e_2\simeq0.03$, $e_3\simeq0.01$ for the observed planetary masses. These numbers are quite close to the ones consistent for the two-body mean motion resonance chain solution found in \cite{2016MNRAS.455L.104G}. Note in passing that their solution is for non-vanishing libration amplitude of the four resonant angles (however their mean values are the same found here for a stable resonant equilibrium).

For completeness, we cite other near-resonant systems of three planets with $k>3$ that are found in the catalogue. The only ones which satisfy our criterion $|\Delta|<0.05$ for both pairs are K2-239 (close to a 3:2 -- 4:3 chain), Kepler-289 (close to a 2:1 -- 2:1 chain), Kepler-226 (close to a 4:3 -- 3:2 chain) and Kepler-431 (close to a 5:4 -- 4:3). Of these, only the latter two satisfy $\Delta>0$ for both pairs.

\section{Conclusions}\label{sec:Conclusions}
In this work, we have generalised the formalism of dissipative divergence of resonant orbits to multi-resonant chains.
The analytical study performed in Section \ref{sec:PlanetaryHamiltonian} allows us to predict the orbital configurations of systems of planets deep in a chain of first order mean motion resonances, and therefore, even though at a lesser degree of precision, of systems that are in resonance with a finite amplitude of libration. Then, we showed in Section \ref{subsec:ResonantRepulsionForThreePlanets} that under the effect of slow dissipation a nearly-resonant system is expected to follow the loci of the equilibrium points of the resonant Hamiltonian \eqref{eq:ReducedHamiltonianWithExpandedKeplerianPart} maintaining the amplitude of libration in an adiabatic manner. Therefore, if the orbital architecture of a system is found near one of these equilibrium points, it is strongly suggestive that the envisioned scenario of slow convergent orbital migration leading to capture into resonance and subsequent orbital divergence due to dissipative evolution really occurred for the system.
In the light of the results presented above, we can draw the following conclusions. 

On the one hand, we must face the fact that the orbital architecture of a significant fraction of the systems of three planets is actually not consistent with these physical mechanisms. More precisely, the majority of the systems are not close to resonance, implying that either they never captured in resonance in the first place, or they escaped from resonance due to a violent instability \citep{2017MNRAS.470.1750I} losing any memory of their resonant dynamical past. 
To ponder these two possibilities, consider first of all that some form of orbital transport \emph{is} expected to take place: for example, the majority of planets with $R_{\mathrm{pl}} \gtrsim 1.6 R_\oplus$ have H/He gaseous atmospheres that cannot be explained by production of volatiles after the formation of the planet \citep{2015ApJ...801...41R}, implying that these planets formed while the protoplanetary disk was still present. The associated planet-disk interaction would then force the planets' orbital elements to change, i.e.\ the planets to migrate. However, orbital migration may not be convergent (\citealt{2015MNRAS.453.1632M,2016MNRAS.458.2051M}), that is, not leading to resonant capture. Moreover, some mechanisms have been proposed to inhibit the capture even in the case of convergent migration, such as turbulence in the disk or $e$-dependent migration rates. Nevertheless, these processes alone do not adequately explain the lack of resonance in the exoplanet sample (e.g.\ \citealt{2017AJ....153..120B,2015ApJ...810..119D,2018MNRAS.481.1538X}). 
For these reasons, it is more likely that capture into mean motion resonance is a common outcome of the early epochs of disk-planet interaction, but the subsequent evolution after the disk removal is subject to instabilities which break the compact configuration. This approach seems to be able to reproduce the observed distribution of period ratios if these instabilities are extremely common \citep{2017MNRAS.470.1750I}.
The primary mechanism through which planets are ejected from resonance, however, remains elusive, and a topic of active research.

On the other hand, systems with orbital properties that are compatible with a (near) resonant state do exist in the exoplanet census.
These include the already known examples mentioned above of Trappist-1, Kepler-223, GJ867 and Kepler-60, and, potentially,  some of the systems analysed in this paper. 
That is, while it is difficult to prove definitively that a given system is \emph{now} in resonance in a formal sense (the resonant angles are in libration), in this work we have developed a method to quantitatively test the consistency of a given orbital architecture with a dynamical history characterised by resonant capture and subsequent dissipative evolution. 
This is achieved through the calculation of the quantity $\bar\delta (a_3/a_2)$ defined in \eqref{eq:WeightedDifferenceOfSmaRatio}, which is obtained directly from the observed semi-major axis ratios and, as we have shown, depends very weakly on the mass ratios between the planets, making the observational uncertainties on the latter quantities irrelevant. Then, using the approach illustrated in Figure \ref{fig:StatDistrOfdeltaa3overa2}, this ``indicator'' can be turned into a quantitative probability that the system is related to the considered resonant chain.
In this sense, we have found that there is a $\sim15\%$, $\sim20\%$ and $\sim80\%$ probability that Kepler-305, Kepler-31 and YZ Cet respectively find themselves this close to resonance merely by chance. Multiplying these probabilities we find that there is only a $\sim2\%$ probability that all three of these systems lie close to resonance just by chance.
This suggests strongly that at least some of them should have had a resonant dynamical history. 
Although the sample is clearly too small to make any meaningful inference, the probabilities of resonant association that we have found indicate that between 1/3 and 2/3 of the systems with $0<\Delta<0.05$ show memory of the processes of resonant capture; this is consistent with the histogram of Figure \ref{fig:DeltDistr}, where the peak wide of the resonance is about 2 times higher than the underlying random-like flat distribution.
 
The architecture of many planetary systems observed by transit is not well constrained by observations. Opportunities for more extensive characterisation will come from missions such as the Transiting Exoplanet Survey Satellite (TESS) or the PLAnetary Transits and Oscillations of stars (PLATO), which are designed to target bright stars to allow for follow-up via further ground-based and space-based observations (with methods such as radial velocity). This will allow for a better quantification of planetary masses, radii, ages of the systems and eccentricities. In the light of this augmented perception that we can expect to acquire, our study outlines the groundwork for further dynamical characterisation of the physical processes that shaped the present-day architectures of extrasolar planetary systems.

\newcommand{\amp}{AMP}

\begin{appendix}
\section{Reduced Hamiltonian to a common planetary mass factor}\label{appendix:ReducedHamiltonian}
In the course of the paper we make implicit use of a reduced Hamiltonian which incorporates the planetary masses through a common planet-to-star mass factor $\tilde m$. In this appendix, we detail the construction of this Hamiltonian and its use in the paper.

Consider three planets, whose physical parameters are labeled 1, 2 and 3 for the inner, middle and outer planet respectively, orbiting around a star of mass $M_*$ on the same plane. Suppose that the planets are (close to) a chain of mean-motion resonance, with nominal semi-major axes $\bar a$ and that the deviations of the semi-major axes from the nominal values are small, and assume that the eccentricities are small enough, so that an analysis to first order in $e$ is valid. These are the working assumptions throughout Section \ref{sec:PlanetaryHamiltonian}.
Having fixed the planet-planet mass ratios $m_1/m_2=\beta_1$ and $m_2/m_3=\beta_2$, we introduce the average planet-star mass ratio
\begin{equation}
\tilde m = \frac{m_1+m_2+m_3}{3 M_*} = \frac{m_1(1+\beta_1^{-1}+\beta_1^{-1}\beta_2^{-1})}{3M_*}.
\end{equation}
Inverting this expression we easily get all planetary masses in terms of $\tilde m$,
\begin{equation}
\begin{split}
m_1=c_1 \tilde m : = \frac{3\beta_1\beta_2 M_*}{1+\beta_2 + \beta_1\beta_2} \tilde m,\\
m_2=c_2 \tilde m : = \frac{3\beta_1 M_*}{1+\beta_2 + \beta_1\beta_2} \tilde m,\\
m_3=c_3 \tilde m : = \frac{3 M_*}{1+\beta_2 + \beta_1\beta_2}\tilde m,
\end{split}
\end{equation}
with coefficients $c$ depending on $M_*$, $\beta_1$ and $\beta_2$ only.

We introduce the modified Delaunay action-angle variables $(\LAMBDONA,\GAMMONA,\LAMBDA,\GAMMA)$ as in \eqref{eq:ModifDelaunayVariables}, but we rescale the actions by the common mass factor $\tilde m$: this gives the following definition for the $\LAMBDONA$'s (we maintain the same notation as the non-rescaled actions for simplicity)
\begin{equation}\label{eq:LambdonasRescaledByTildem}
\begin{split}
\LAMBDONA_1=\frac{3\beta_1\beta_2 M_*}{1+\beta_2 + \beta_1\beta_2}\sqrt{\GravC M_* a_1} = c_1 \sqrt{\GravC M_* a_1} ,\\
\LAMBDONA_2=\frac{3\beta_1 M_*}{1+\beta_2 + \beta_1\beta_2}\sqrt{\GravC M_* a_2} = c_2 \sqrt{\GravC M_* a_2},\\
\LAMBDONA_3=\frac{3 M_*}{1+\beta_2 + \beta_1\beta_2}\sqrt{\GravC M_* a_3} = c_3 \sqrt{\GravC M_* a_3},
\end{split}
\end{equation}
and the same formal definition of $\GAMMONA=\LAMBDONA e^2/2$ at lowest order in $e$.
We now introduce the Hamiltonian for the problem, that is the sum of the Keplerian Hamiltonian \eqref{eq:KeplerianPartInModifDelaunayVariables} and the resonant interaction Hamiltonian \eqref{eq:ResonantHamiltonianInModifDelaunayVariables} to first order in $e$. 
However since we have rescaled the actions by $\tilde m$, in order for the Hamilton equations to be conserved we must also rescale the Hamiltonian itself by $\tilde m$.
As in Section \ref{sec:PlanetaryHamiltonian}, since we are not considering large deviations in the semi-major axes from their nominal values, and since the resonant Hamiltonian is already of order $\mathcal O(e)$, we evaluate the resonant Hamiltonian on the nominal values $\bar\LAMBDONA$ defined from $\bar a$ using \eqref{eq:LambdonasRescaledByTildem}.
It is then easy to see that the rescaled Keplerian Hamiltonian takes the form
\begin{equation}
\Ha_{kepl} = -\sum_{i=1}^3 \frac{c_i^3}{2}\left(\frac{\GravC M_*}{\LAMBDONA_i}\right)^2,
\end{equation} 
and is therefore independent of $\tilde m$, while the rescaled resonant part will have a multiplicative coefficient $\tilde m$:
\begin{equation}
\begin{split}
\Ha_{res}		&= \tilde m \left[		-\frac{\GravC^2 M_* c_1 c_2^3}{\bar\LAMBDONA_2^2} \right. \\
				&\quad\quad \times\left(f_{res}^{(1,in)}\sqrt{\frac{2\GAMMONA_1}{\bar\LAMBDONA_1}} 
				\cos\big(k^{in}\LAMBDA_2-(k^{in}-1)\LAMBDA_1 + \GAMMA_1\big)\right.\\
				&\quad\quad\quad \left.+ f_{res}^{(2,in)}\sqrt{\frac{2\GAMMONA_2}{\bar\LAMBDONA_2}} 
				\cos\big(k^{in}\LAMBDA_2-(k^{in}-1)\LAMBDA_1 + \GAMMA_2\big)\right) + \\
			&\quad 	-\frac{\GravC^2 M_* c_2 c_3^3}{\bar\LAMBDONA_3^2} \\
				&\quad\quad \times\left(f_{res}^{(1,out)}\sqrt{\frac{2\GAMMONA_2}{\bar\LAMBDONA_2}} 
				\cos\big(k^{out}\LAMBDA_3-(k^{out}-1)\LAMBDA_2 + \GAMMA_2\big)\right.\\
			&\quad\quad\quad \left.\left.+ f_{res}^{(2,out)}\sqrt{\frac{2\GAMMONA_3}{\bar\LAMBDONA_3}} 
				\cos\big(k^{out}\LAMBDA_3-(k^{out}-1)\LAMBDA_2 + \GAMMA_3\big)\right)\right];
\end{split}
\end{equation}
From here, the sequence of changes of variables detailed in Section \ref{sec:PlanetaryHamiltonian} can be performed using the same formal definitions for the new rescaled variables.

Already from the Hamiltonian written in terms of the rescaled variables $\LAMBDONA$ and $\GAMMONA\propto e^2$ one can see the following. Assuming a fixed equilibrium value of the semi-major axes (that is, of the $\LAMBDONA$'s), a change in the planet-star mass factor $\tilde m$ will have the only effect to rescale the equilibrium values of all $\sqrt{\GAMMONA}\propto e$ by the same quantity. In this configuration, the equilibria of the semi-major axis ratios $(a_2/a_1)_{eq}$ and $(a_3/a_2)_{eq}$ remain independent of $\tilde m$.
\end{appendix}

\end{document}